\documentclass[12pt]{article}

\usepackage{amssymb,amsmath,amsfonts,bbm,eurosym,geometry,ulem,graphicx,caption,color,setspace,sectsty,comment,footmisc,caption,natbib,pdflscape,array,subfiles}

\usepackage[FIGTOPCAP]{subfigure}

\usepackage{abstract}

\usepackage[colorlinks=true,linkcolor=blue, allcolors=blue]{hyperref}
\usepackage{threeparttable}
\usepackage[toc,page]{appendix}
\usepackage{pdflscape}
\usepackage{afterpage}
\usepackage{changepage}
\usepackage{booktabs}
\hypersetup{breaklinks=true}
\usepackage{lscape} 

\newcolumntype{L}[1]{>{\raggedright\let\newline\\arraybackslash\hspace{0pt}}m{#1}}
\newcolumntype{C}[1]{>{\centering\let\newline\\arraybackslash\hspace{0pt}}m{#1}}
\newcolumntype{R}[1]{>{\raggedleft\let\newline\\arraybackslash\hspace{0pt}}m{#1}}

\setcitestyle{authoryear,open={(},close={)}}

\begin{document}
\title{\vspace{-1.7cm}The geographic spread of COVID-19 correlates with the structure of social networks as measured by Facebook\thanks{Date: \today. Public versions of the social connectedness data used in this paper, as well as similar data for a wide range of other geographies, are accessible at \url{https://data.humdata.org/dataset/social-connectedness-index}. The full replication code is available at \url{https://github.com/social-connectedness-index/example-scripts}. The authors have a research consulting relationship with Facebook. Since this project only uses data that is available to the broader research community, nobody at Facebook reviewed the contents of this paper.}}
\author{\hspace{-0.3cm}
	Theresa Kuchler\thanks{New York University, Stern School of Business. Email: \href{mailto:tkuchler@stern.nyu.edu}{tkuchler@stern.nyu.edu}} \and
	\hspace{-0.3cm}Dominic Russel\thanks{New York University, Stern School of Business. Email: \href{mailto:drussel@stern.nyu.edu}{drussel@stern.nyu.edu}}
	\and \hspace{-0.3cm}
	Johannes Stroebel\thanks{New York University, Stern School of Business. Email: \href{mailto:jcstroebel@gmail.com}{johannes.stroebel@nyu.edu} (Corresponding)}\vspace{0.22cm} 
		}
\date{}
\maketitle

\begin{abstract}\noindent We use aggregated data from Facebook to show that COVID-19 is more likely to spread between regions with stronger social network connections.  Areas with more social ties to two early COVID-19 “hotspots” (Westchester County, NY, in the U.S. and Lodi province in Italy) generally had more confirmed COVID-19 cases by the end of March. These relationships hold after controlling for geographic distance to the hotspots as well as the population density and demographics of the regions. As the pandemic progressed in the U.S., a county's social proximity to recent COVID-19 cases and deaths predicts future outbreaks over and above physical proximity and demographics. In part due to its broad coverage, social connectedness data provides additional predictive power to measures based on smartphone location or online search data. These results suggest that data from online social networks can be useful to epidemiologists and others hoping to forecast the spread of communicable diseases such as COVID-19.\bigskip
\end{abstract}

\noindent To forecast the geographic spread of communicable diseases such as COVID-19, it is valuable to know which individuals are likely to physically interact \citep[][]{y2018charting}. In particular, since social ties shape patterns of physical interaction, observing the strength of social connections between cities and regions can be useful for determining a locality’s risk of future disease outbreaks. Yet, the geographic structure of social networks is usually difficult to measure on a national or global scale. In this paper, we overcome this challenge by using aggregated data from Facebook to measure social connections between regions. We then show that these connectedness measures can help forcecast the geographic spread of communicable diseases such as COVID-19. 

We construct a measure of the social connectedness between U.S. counties and between Italian provinces. This \emph{Social Connectedness Index} captures the probability that Facebook users in a pair of these regions are Facebook friends with each other \citep{Bailey2018_measure}. We hypothesize that regions connected through many friendship links are likely to have more physical interactions between their residents, providing opportunities for the spread of communicable diseases. Indeed, our measure has been shown to be predictive of travel patterns across Europe \citep{bailey2020_euro} and within urban areas \citep{bailey2020_urban}, suggesting it contains important information about real-world interactions. Most directly, \cite{coven2020urban} use our \emph{Social Connectedness Index}  to show that counties with higher levels of social connectedness to New York City were more likely to be destinations for those fleeing the city during the pandemic, providing direct evidence for our propopsed mechanism.

After introducing our \emph{Social Connectedness Index}, we show that regions with stronger social ties to early COVID-19 “hotspots” — Westchester County, NY, in the U.S., and Lodi province in Italy — had more documented COVID-19 cases per resident as of March 30, 2020. These relationships are robust to controlling for the geographic distance to these early hotspots, as well as demographic characteristics of the regions. Social connectedness to Westchester has more predictive power for forecasting county-level COVID-19 cases than social connectedness to any other county outside the New York-Newark CSA. These case studies provide initial evidence that social connectedness might serve as a valuable predictive measure in addition to physical distance and other inputs to current epidemiological models.

We then exploit the changing geography of the pandemic in the U.S. to conduct a more systematic in-sample analysis. We construct regional measures of COVID-19 exposure through social connections (``social proximity to cases”) and physical distance (``physical proximity to cases”). We find that changes in a county's social proximity to cases in one time period are strongly correlated with the county's subsequent growth in own local cases. Even after controlling for physical proximity to cases and other regional demographics, a doubling in social proximity to cases in one two-week period corresponds to a 24.9\% increase in own cases in the next two-week period. These results are unlikely to be explained by differential testing between regions, as an increase in social proximity to deaths in one period also corresponds to an increase in actual deaths in the next period.

To mimic a real-world epidemiological use case, we also conduct a simple out-of-sample prediction exercise. We find that models that include our measure of social proximity to cases are better able to predict a region's future case growth than alternative models that rely only on geographic distance and other demographics. We also compare the predictive value of social proximity to cases to measures from Google searches related to COVID-19 symptoms and the smartphone-based Location Exposure Index (LEX) introduced by \cite{couture2020measuring}. In counties with both LEX and Google search data, social proximity to cases provides only small additional predictive value --- perhaps not surprisingly, given that the real-world movement of people captured by the LEX is precisely the mechanism we conjecture explains the predictive power of social proximity to cases. However, when using the best available model to make predictions for \emph{all} U.S. counties (for many of which no LEX data or Google search information is available), models that include social proximity to cases sizably improve accuracy. This highlights one important advantage of social connectedness data: its broad coverage and global availability.

Our use of the \emph{Social Connectedness Index} to forecast COVID-19 spread adds to an active body of research that studies how aspects of social media and internet-usage patterns can be used for tracking and preventing disease \citep[for an overview, see][]{aiello2020_social}. One strand of this literature uses the content of individuals' internet searches or social media posts; most famously, Google Flu Trends used search queries related to influenza for early outbreak detection \citep{ginsberg2009detecting}. Other researchers have also used content from Twitter posts \citep{rodriguez2018twitter, jahanbin2020using}, Facebook likes \citep{gittelman2015new}, Wikipedia searches \citep{generous2014global}, and Instagram posts \citep{correia2016monitoring} to predict public health outcomes. A second strand of research, which has received much attention during the COVID-19 pandemic, uses geolocation data to track individuals' movement patterns. These data have been used to explore the determinants and effects of social distancing behavior \cite[for an overview, see][]{vox_social_distancing_lit_review}, as well as forecast disease spread \citep[e.g.,][]{jia2020population, bengtsson2015using, wesolowski2012quantifying, wesolowski2015impact, peixoto2020modeling}. A third strand of that work uses crowdsourced information, including surveys, to monitor disease symptoms and detect outbreaks \cite[see][]{facebook_survey, smolinski2015flu, paolotti2014web}.

In comparison to this literature, our stable network-based measure is less likely to suffer from changes in internet behavior or seasonality, both of which have hampered Google Flu Trends \citep{olson2013reassessing}. In addition, our measures do not require individuals to have experienced symptoms, which potentially allows us to identify at-risk localities  before disease transmission.\footnote{However, this suggests that our data might partner well with these measures. For example, if one can detect an early outbreak using surveys, they could then predict (and potentially prevent) the next outbreak using information on social connectedness.} Finally, because our measures are based only on aggregated connections (instead of individual movement), they are easily accessible to researchers and consistently available for a large number of granular geographies around the world. For example, the \emph{Social Connectedness Index} is available at the NUTS3 level in Europe, the GADM2 level in the Indian Subcontinent and Canada, and the GADM1 level throughout much of the rest of the world.\footnote{Interested researchers may also access U.S. ZCTA-level data by emailing \href{mailto:sci_data@fb.com}{sci\_data@fb.com}.} The index not only measures connections \emph{within} countries, but also \emph{between} countries, which may be otherwise challenging with mobility data from different cellphone providers (and important for tracking the international spread of communicable diseases).

More generally, our results add to a literature that has applied aspects of network theory to build spatial epidemiological models \citep[for overviews, see][]{keeling2005networks, keeling2011modeling, danon2011networks}. Works in this literature move beyond the basic assumption that individuals within a population are “fully mixed,” or equally likely to interact; instead, they better represent the dynamics of real-world connections \citep[e.g.,][]{newman2002spread, klovdahl1985social, klovdahl1994social, mossong2008social, yang2020quantifying}. While some of these studies parameterize models with information on local networks, we are unaware of any that introduce a measure with comparably high levels of coverage and granularity. Our hope is that our unique measure of social connectedness can help parameterize future epidemiological work. In addition, we hope that the \textit{Social Connectedness Index} can advance the literature on the determinants and effects of urban and regional social networks \citep[see][]{bailey2020_urban, kim2017interactions, buchel2020socint, mossay2011spatial, brueckner2008sprawl, glaeser1992growth}.

It is important to note that our objective in this paper is not to incorporate social connectedness into a state-of-the-art epidemiological model. Instead, we provide a unique measure to assess regions' outbreak risk, answering the call of \cite{avery2020policy}, among others, who highlight an ``urgent need'' for ``creative and entrepreneurial methods" of interpreting and sharing data to model coronavirus spread. To that end, the data used in this paper, as well as similar data for a number of other geographies, are available at \url{https://data.humdata.org/dataset/social-connectedness-index}. We encourage interested researchers to use them.

\section{Data Description} \label{sec:data}

To measure the intensity of social connectedness between locations, we use a de-identified and aggregated snapshot of all active Facebook users and their friendship networks from March 2020.\footnote{We use the data from March 2020, since this this allows our analyses to correspond most closely to a real-time forecasting exercise. The publicly available \textit{Social Connectedness Index} data is based on a snapshot from August 2020. Since the SCI is extremely stable over time, results do not change across the various data sets.} As of the end of 2019, Facebook had nearly 2.5 billion monthly active users around the world: 248 million in the U.S. and Canada, 394 million in Europe, 1.04 billion in Asia-Pacific, and 817 million in the rest of the world \citep{fb_10k}. The data therefore has extremely wide coverage, and provides a unique opportunity to map the geographic structure of social networks around the world. Locations are assigned to users based on their information and activity on Facebook, including their public profile information, and device and connection information. Establishing a connection on Facebook requires the consent of both individuals, and there is an upper limit of 5,000 on the number of connections a person can have. As a result, Facebook connections are generally more likely to be between real-world acquaintances than links on many other social networking platforms. 

Our measure of the social connectedness between two locations $i$ and $j$ is the \textit{Social Connectedness Index (SCI)} introduced by  \cite{Bailey2018_measure}:
\begin{equation}
    \label{eq:sci}
    Social \ Connectedness_{i,j} = \frac{FB \ Connections_{i,j}}{FB \ Users_i*FB \ Users_j}.
\end{equation}
Here, $FB \ Connections_{i,j}$ is the total number of Facebook friendship links between Facebook users living in location $i$ and Facebook users living in location $j$. $FB \ Users_i$ and $FB \ Users_j$ are the number of active users in each location. $Social \ Connectedness_{i,j}$ thus measures the relative probability of a Facebook friendship link between a given Facebook user in location $i$ and a given Facebook user in location $j$: if this measure is twice as large, a given Facebook user in region $i$ is twice as likely to be friends with a given Facebook user in region $j$. 

In previous work, we have shown that this measure predicts a large number of important economic and social interactions. For example, social connectedness as measured through Facebook friendship links is strongly related to patterns of sub-national and international trade \citep{bailey2020_international}, patent citations \citep{Bailey2018_measure}, and investment decisions \citep{kuchlerinstitution2020}.\footnote{A growing body of research also uses this measure to study issues related to COVID-19. For example, \cite{bailey2020_distancing} use the $SCI$, along with individual-level data from Facebook, to show that social network exposure to COVID-19 cases shapes individuals’ social distancing  behaviors. \cite{Holtz19837} use $SCI$ data to document spillover effects in state-level COVID-19 health policies. \cite{charoenwong2020social} and \cite{makridis2020learning} work with the $SCI$ to study other behavioral effects of exposure to COVID-19 through friends.} More generally, we have found that information on individuals' Facebook friendship links can help understand their product adoption decisions  and their housing and mortgage choices \citep{bailey2018_economic,bailey2019house, bailey2019_peer}.

Data on COVID-19 cases in the U.S. by county come from \href{https://github.com/CSSEGISandData/COVID-19}{Johns Hopkins University Center for Systems Science and Engineering}. Data on COVID-19 cases for each Italian province come from the \href{https://github.com/pcm-dpc}{Italian Dipartimeno della Protezione Civile}. Because differential testing across regions may introduce bias in case-based results, we will also use information on COVID-19 related deaths from each source in Section \ref{sec:series}.

\section{Early Hotspot Analysis} \label{sec:early}

In this section, we explore how the domestic spread of confirmed COVID-19 cases is related to the social connectedness to two early COVID-19 ``hotspots'': Westchester County, NY, in the U.S., and Lodi Province in Italy. Westchester County includes New Rochelle, a community that had the first major confirmed COVID-19 outbreak in the eastern United States \citep{chappell_npr}. By March 20th, the county had over 9,300 cases, second only to nearby New York City. Additionally, a number of articles reported that wealthy residents from Westchester and the New York area had fled to other parts of the U.S. \citep{tully_nyt}, providing a vector that could potentially spread the disease. Indeed, geneticists and epidemiologists later found that travel from New York seeded much of the first wave of U.S. COVID-19 outbreaks \citep{carey_nyt}.  Social connections to Westchester may thus provide particularly important information for tracking early COVID-19 spread, especially given that \cite{coven2020urban} find that social connectedness to New York City predicted travel patterns from the city early in the pandemic. Lodi is an Italian province of around 230,000 inhabitants in the heavily impacted region of Lombardy. It contains Codogno, where the earliest cases of COVID-19 in Italy were detected, and was at the center of Italy's outbreak \citep{horowitz_nyt}.

\begin{figure}[p!]
  \caption{Social Network Distributions from Westchester and COVID-19 Cases in the U.S.}
  \label{fig:usa}
  \subfigure[Log of SCI to Westchester County, NY]{
    \includegraphics[clip, trim=0cm 0cm 0cm 0cm, width=0.5\linewidth]{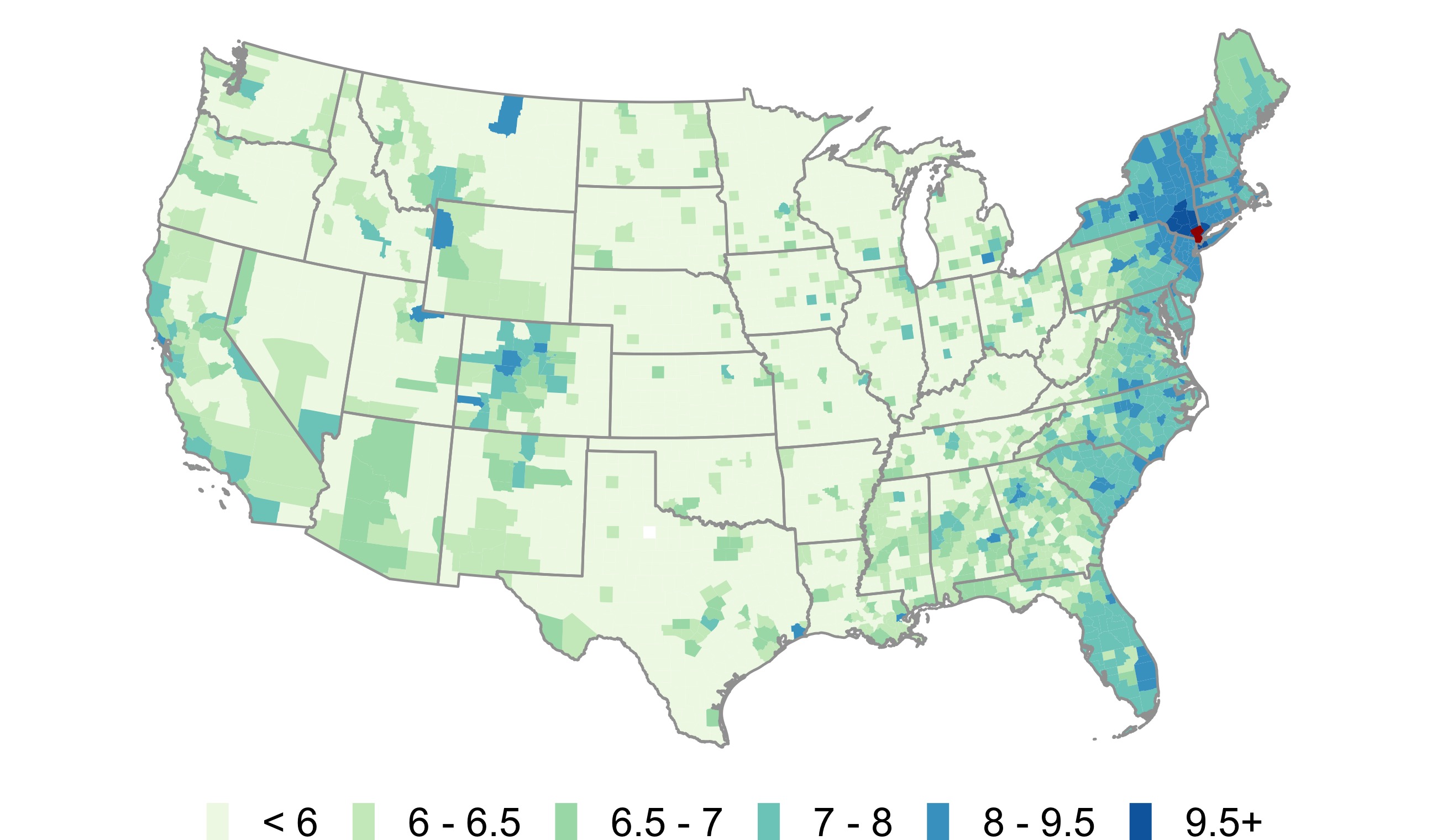}}
  \subfigure[COVID-19 Cases per 10k Residents by County]{
    \includegraphics[clip, trim=0cm -5cm 0cm 0cm, width=0.5\linewidth]{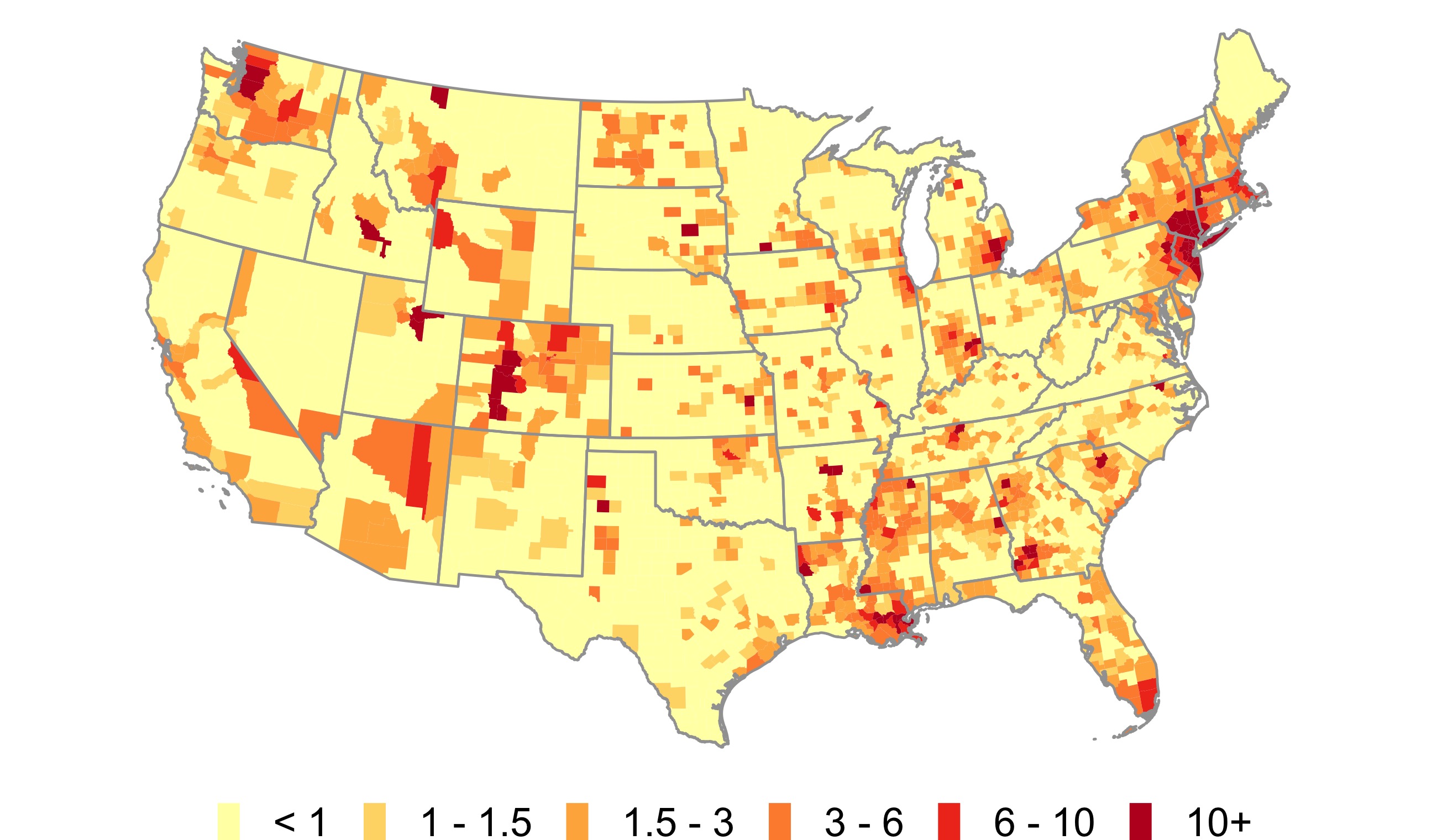}}
 \subfigure[Westchester binscatter without controls]{
      \includegraphics[clip, trim=0.25cm 0.25cm 0.25cm 0.25cm, width=0.5\linewidth]{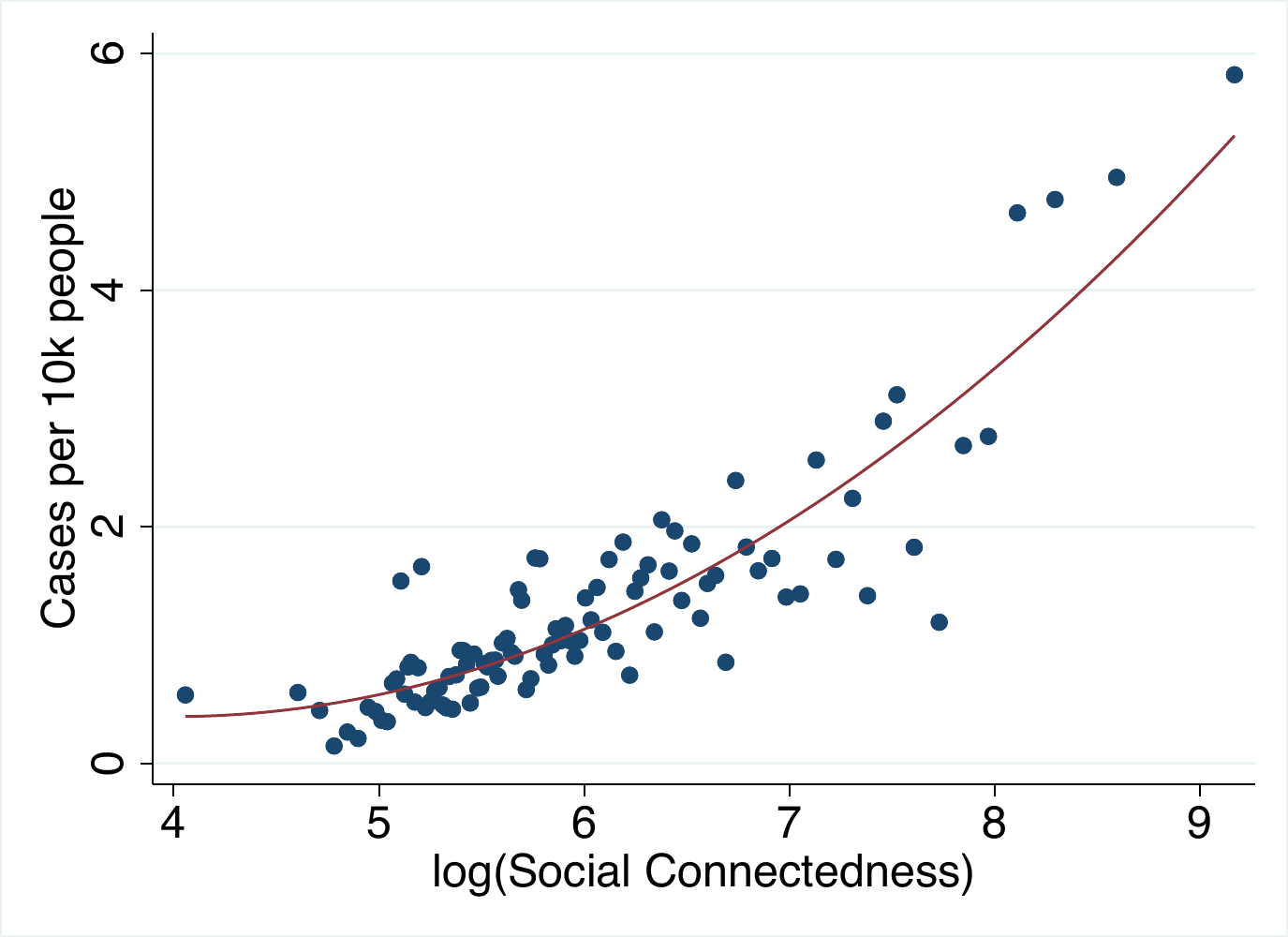}}
 \subfigure[Westchester binscatter with controls]{
      \includegraphics[clip, trim=0.25cm 0.25cm 0.25cm 0.25cm, width=0.5\linewidth]{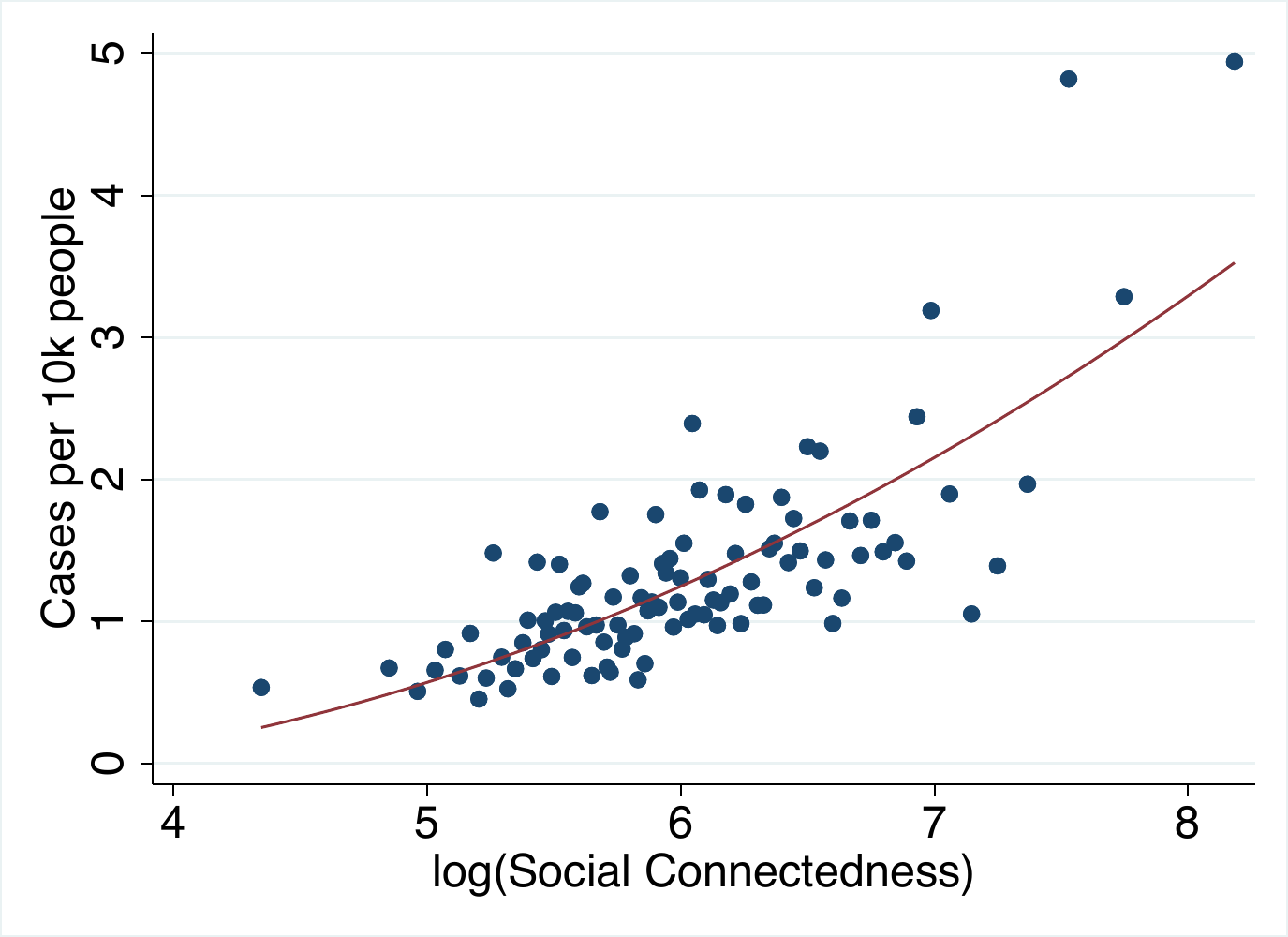}}
      	\begin{minipage}{\textwidth} \setstretch{.9} \medskip
\footnotesize{{\bf Note:} Panel (a) shows the social connectedness to Westchester for U.S. counties. Panel (b) shows the number of confirmed COVID-19 cases per 10,000 residents by U.S. county on March 30, 2020. Panels (c) and (d) show binscatter plots with counties more than 50 miles from Westchester as the unit of observation. To generate the plot in Panel (c), we group $log(SCI)$ into 100 equal-sized bins and plot the average against the corresponding average case density. Panel (d) is constructed in a similar manner. However, we first regress $log(SCI)$ and cases per 10,000 residents on a set of control variables and plot the residualized values on each axis. Red lines show quadratic fit regressions. The controls for Panel (d) are 100 dummies for the percentile of the county's geographic distance to Westchester; population density; median household income; and dummies for the six National Center for Health Statistics Urban-Rural county classifications.}
\end{minipage}
\end{figure}

Panel (a) of Figure \ref{fig:usa} shows a heatmap of the social connectedness of Westchester County, NY, to other U.S. counties; darker colors correspond to stronger social ties. Panel (b) shows the distribution of COVID-19 cases per 10,000 residents across U.S. counties on March 30, 2020, with darker colors corresponding to higher COVID-19 prevalence. These maps show a number of similarities. Perhaps most notably, coastal regions and urban centers appear to have both high levels of connectedness to Westchester and larger numbers of COVID-19 cases per resident. But a number of more subtle patterns also emerge. Both measures are high in the communities along the Florida coast (in particular along the southeastern coast, near Miami), in western and central Colorado (in particular in areas with ski resorts), and in the upper Northeast. These areas are all popular vacation destinations and second home locations for many well-heeled residents of Westchester. Indeed, the governors of Florida and Rhode Island publicly lamented the number of New York area residents fleeing to their states and spreading COVID-19 \citep{mower_tbt, carlisle_time}. By contrast, many areas that are geographically closer but less socially connected to Westchester, such as counties in western Pennsylvania and West Virginia, had fewer confirmed COVID-19 cases on March 30. There are also a number of patterns of COVID-19 prevalence that connectedness to Westchester alone cannot explain. Areas around King County, WA (Seattle), for example, have relatively low connectedness to Westchester, but were an independent early hotspot of COVID-19.

The two bottom panels of Figure \ref{fig:usa} explore the relationship between COVID-19 prevalence and social ties to Westchester more formally. Panel (c) shows a binscatter plot of social connectedness to Westchester County and the number of COVID-19 cases per 10,000 residents. We exclude those counties within 50 miles of Westchester County: while those areas have strong social links to Westchester, they are also close enough geographically such that their populations might interact physically with Westchester residents even in the absence of social links (e.g., in supermarkets and houses of worship). There is a strong positive relationship between social ties to Westchester and COVID-19 prevalence. Quantitatively, a doubling of a county's social connectedness to Westchester is associated with an increase of about 0.88 COVID-19 cases per 10,000 residents. The R-squared of this relationship is 0.093, suggesting that, in a statistical sense, 9.3\% of the cross-county variation in COVID-19 cases can be explained by counties' social connectedness to Westchester.

One concern with interpreting these initial correlations is that they might be primarily picking up other factors that affect the spread of COVID-19, and that are correlated with social connectedness to Westchester. Specifically, even after dropping counties within 50 miles of Westchester, the correlations might be primarily picking up geographic distance to Westchester (which is related to the number of friendship links to Westchester). As a result, including social connectedness might not improve predictive power for models that already control for some of these other variables. In Panel (d), we therefore present a binscatter plot of the relationship between social connectedness to Westchester County and COVID-19 cases that controls for a number of these possible confounding variables (in addition to excluding nearby counties). Most importantly, we non-parametrically control for the geographic distance between each county and Westchester County by including 100 dummies for percentiles of that distance. We also control for median income, population density, and a classification of how urban/rural a county is. Even conditional on these other factors,  Panel (d) shows a strong positive relationship between COVID-19 cases as of March 30, 2020 and social connectedness to Westchester County. With these controls, a doubling of a county's social connectedness to Westchester is associated with an increase of about 0.80 COVID-19 cases per 10,000 residents. The total R-squared of the statistical relationship is 0.190, while the incremental R-squared from controlling for social connectedness to Westchester is 0.037.

\begin{figure}[htb]
  \caption{Incremental $R^2$ from Adding Connections to Individual U.S. Counties}
  \label{fig:placebo}
    \subfigure[Excluding Counties within 50 Miles]{
    \includegraphics[clip, trim={0.1cm 0cm 0.1cm 0.1cm}, width=0.49\linewidth]{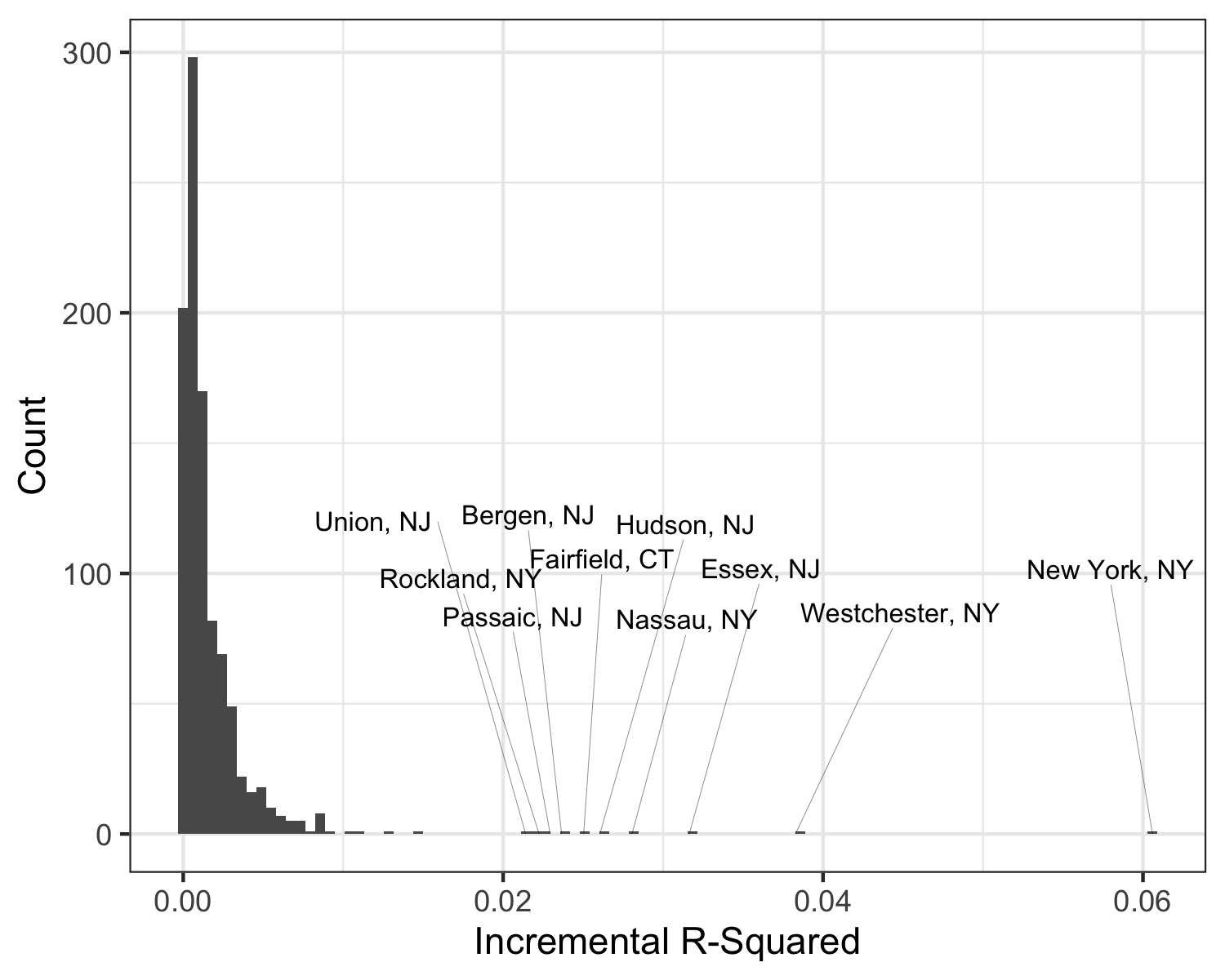}}
  \subfigure[Excluding Counties within 150 Miles]{
    \includegraphics[clip, trim=0.1cm 0cm 0.1cm 0.1cm, width=0.49\linewidth]{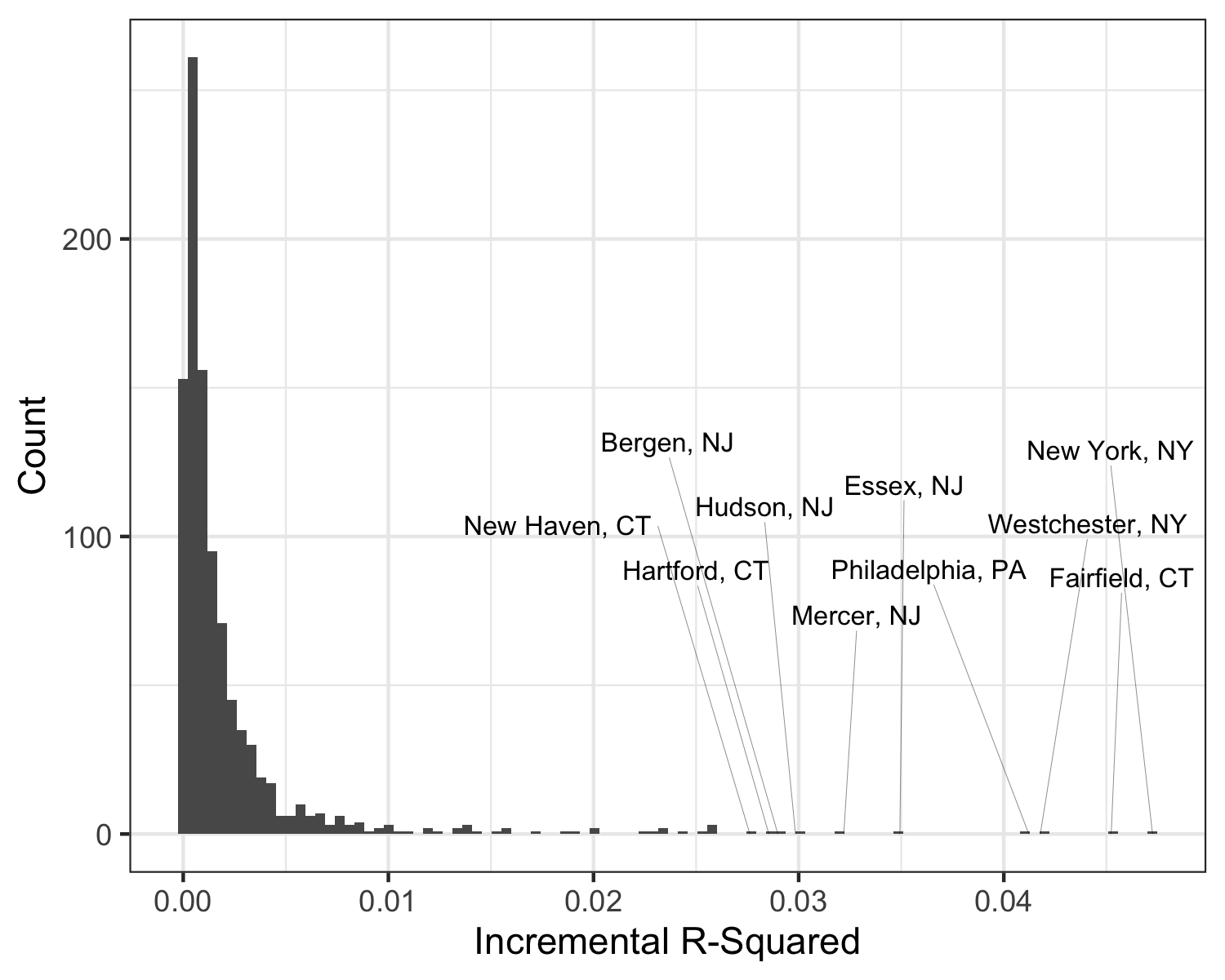}}
     \begin{minipage}{\textwidth} \setstretch{.9} \medskip
\footnotesize{{\bf Note:} Panels show results from regressions to predict COVID-19 cases per 10k people by county on March 30, 2020. The incremental $R^2$ is the increase in $R^2$ from adding $log(SCI)$ and $log(SCI)^2$ to a particular U.S. county, over and above a set of baseline control variables: 100 dummies for percentiles of distance to the county under investigation; population density; median household income; and dummies for the six National Center of Health Statistics Urban-Rural county classifications. The graphs show the distributions over the incremental $R^2$s for adding social connectedness to each county with a population over 50,000  in turn. Each regression in panels (a) and (b) excludes counties within 50 and 150 miles of the county of interest, respectively.  In each panel the 10 largest incremental $R^2$ are labeled.}
\end{minipage}
\end{figure}

Another potential concern stems from the fact that the underlying social network and the site of the initial hotspot are nonrandom. This may confound our interpretation if, for example, counties with ties to Westchester were also destinations for European travelers seeding the virus in the United States. To contextualize the effect of connections to Westchester in particular, we next run ``placebo'' regressions, identical to the one shown for Westchester in panel (d) of Figure \ref{fig:usa}, for every U.S. county with a population over 50,000. Figure \ref{fig:placebo} shows the incremental R-squared from adding social connectedness in each of these regressions. Panel (a) excludes counties within 50 miles of the chosen county, as in Figure \ref{fig:usa}. Westchester's 0.037 incremental R-squared is second only to New York City, and each top 10 county is in the New York-Newark Combined Statistical Area (CSA).\footnote{Although New York City contains multiple counties, early NYC COVID-19 data was not disaggregated. As such, we combine the counties in this section. NYC is disaggregated in all analyses in Section \ref{sec:series}.}  That connections to each of these counties matters so strongly suggests that, although Westchester contained the earliest discovered COVID-19 outbreak in the eastern U.S., community spread may have already been present in many neighboring counties. Panel (b) shows results for regressions excluding counties within 150 miles of the chosen county. Doing so will exclude New York City and Westchester cases from every regression for a New York-Newark CSA county.\footnote{The maximum distance from any county in the CSA to New York City and Westchester is 91 miles and 104 miles, respectively.} Counties within the CSA remain as 9 of the top 10, including the top 3: New York City, Fairfield, and Westchester.\footnote{Anecdotally, \cite{williamson_nyt} linked Fairfield to COVID-19 spread early in the pandemic.} These findings highlight that social connections to other counties that may have similar demographics to Westchester, but that did not have an early COVID-19 outbreak, do not help with forecasting COVID-19 spread. In turn, this suggests our previous results are not due to omitted variables whereby counties with links to Westchester are more susceptible to COVID-19 outbreaks for some other reason. 

It is important to highlight that the purpose of this exercise is to demonstrate the \textit{predictive power} of social connectedness measured via online social networks for COVID-19 prevalence. The control variables highlight that the \textit{Social Connectedness Index} has such predictive power over and above a number of variables on which data is already easily available, and that may partially proxy for social connections in models of communicable disease spread. We will benchmark this predictive power against other measures, such as smartphone location pings and Google searches for COVID-19 symptoms, in Section \ref{sec:series}.

\begin{figure}[p!]
  \caption{Social Network Distributions of Lodi and COVID-19 Cases in Italy}
  \label{fig:italy}
    \subfigure[Percentile of SCI to Lodi Province, Italy]{
    \includegraphics[clip, trim={0.1cm 10cm 0.1cm 0.1cm}, width=0.48\linewidth]{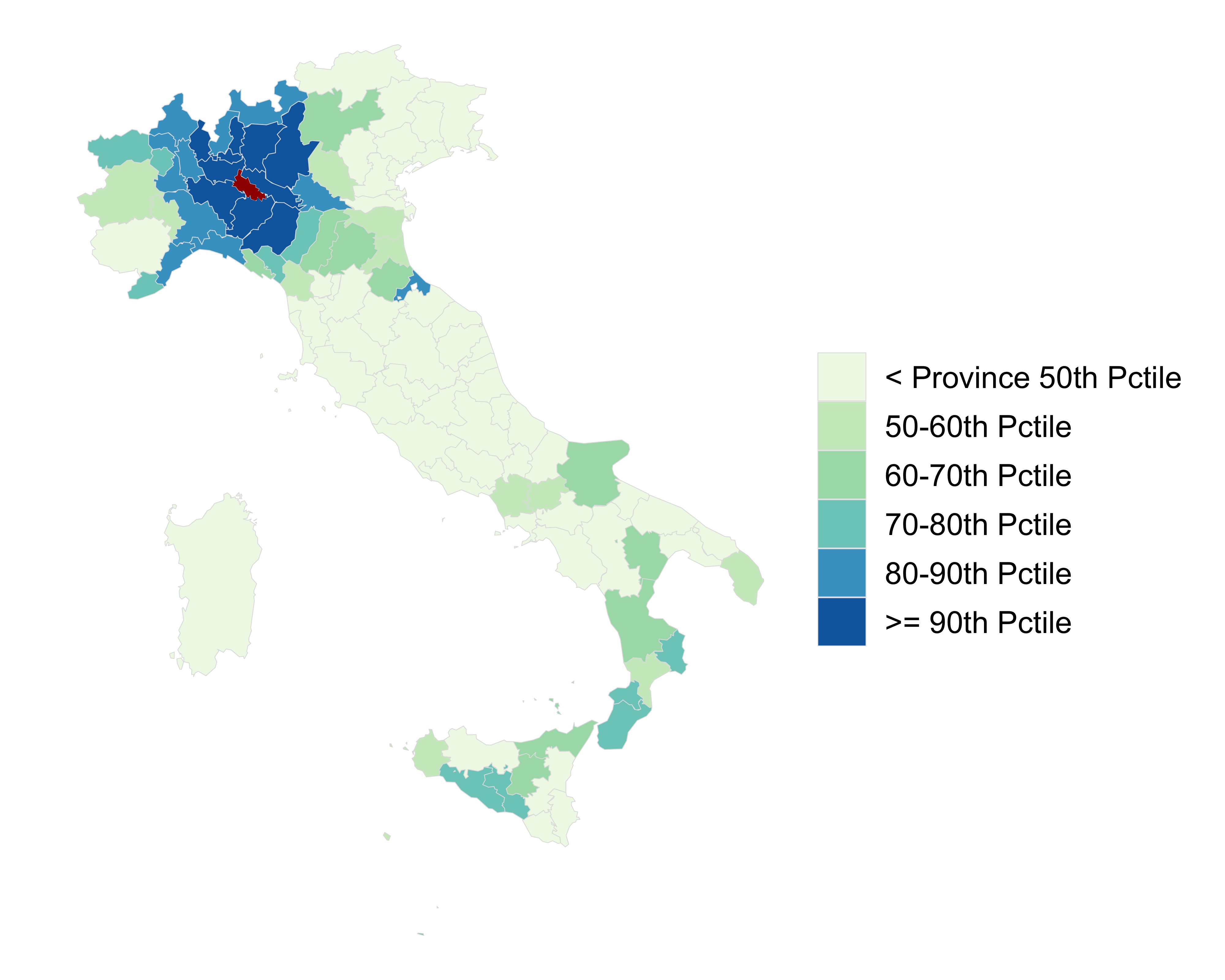}}
  \subfigure[COVID-19 Cases per 10k Residents by Province]{
    \includegraphics[clip, trim=0.1cm 10cm 0.1cm 0.1cm, width=0.48\linewidth]{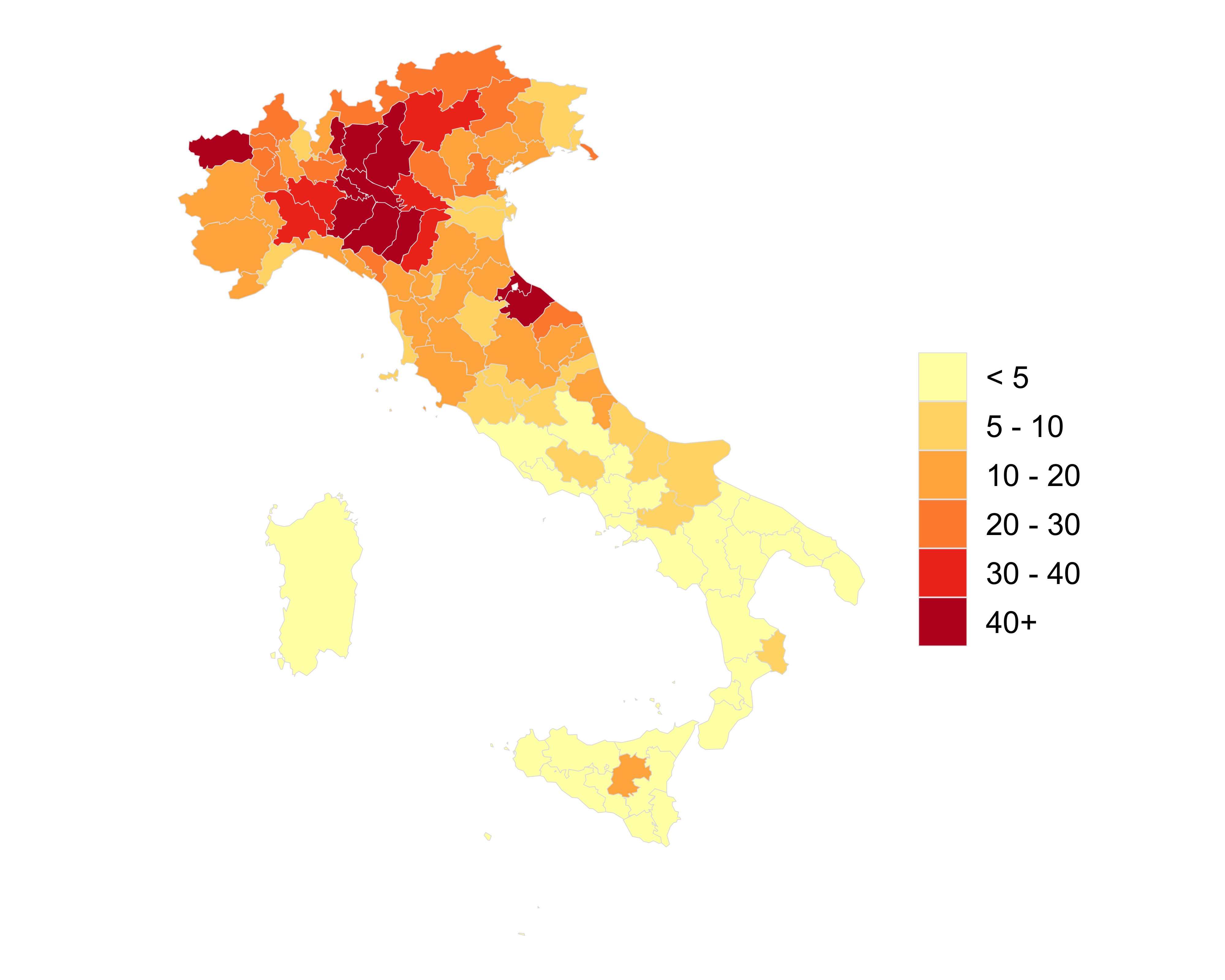}}
 \subfigure[Lodi binscatter without controls]{
      \includegraphics[clip, trim=0.25cm 0.25cm 0.25cm 0.25cm, width=0.5\linewidth]{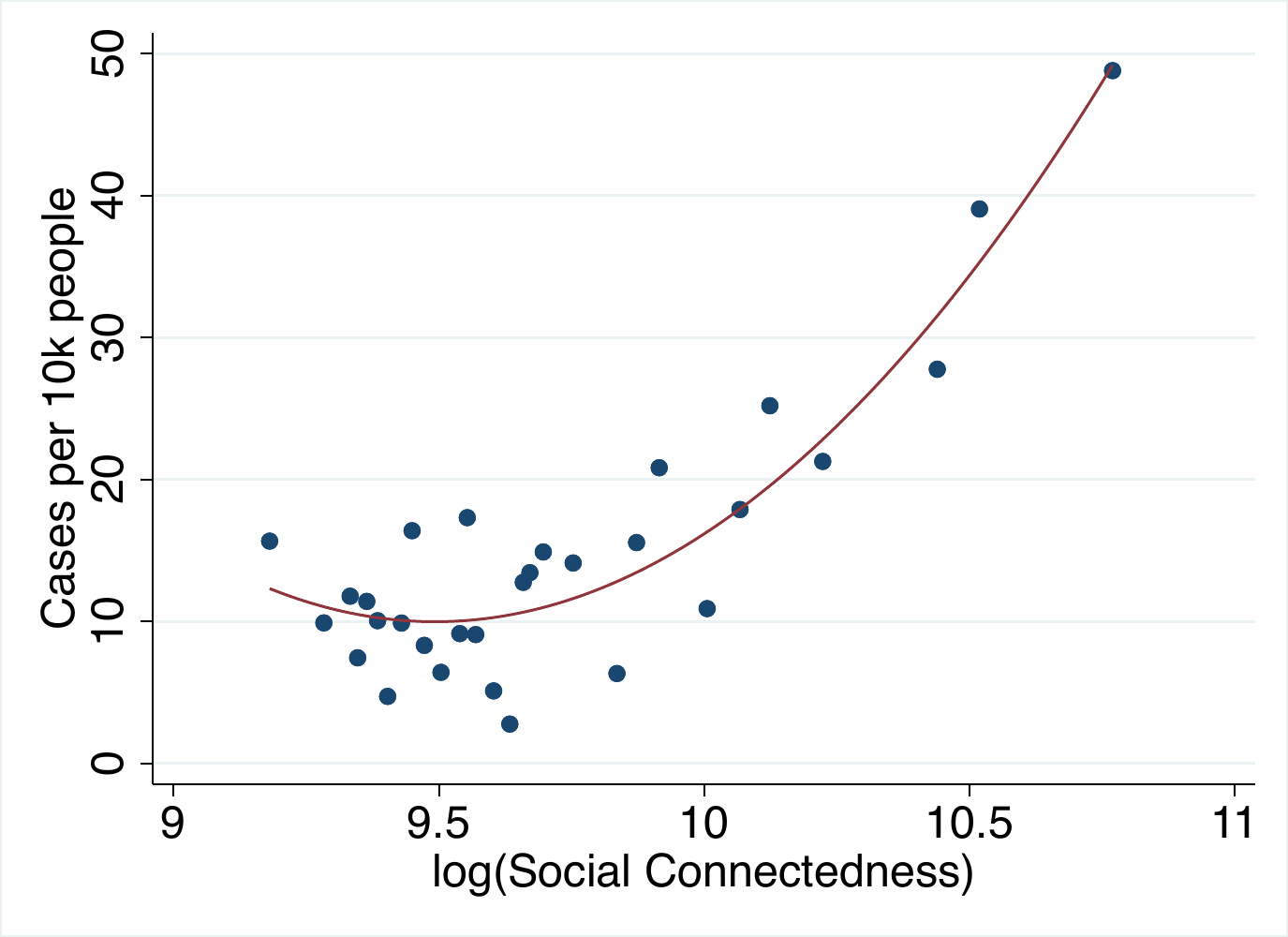}}
   \subfigure[Lodi binscatter with controls]{
      \includegraphics[clip, trim=0.25cm 0.25cm 0.25cm 0.25cm, width=0.5\linewidth]{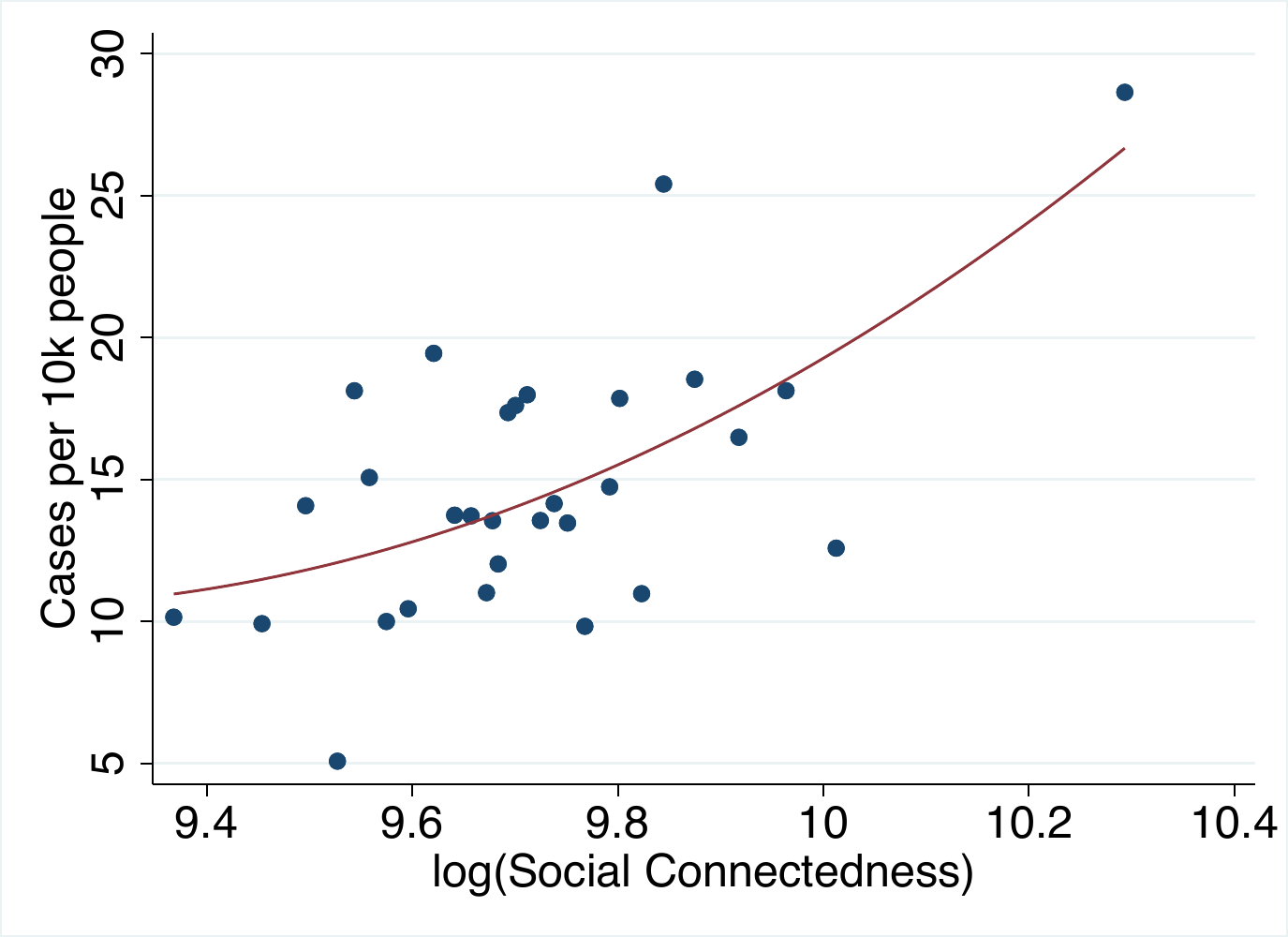}}
      	\begin{minipage}{\textwidth} \setstretch{.9} \medskip
\footnotesize{{\bf Note:} Panel (a) shows the social connectedness to Lodi for Italian provinces. Panel (b) shows the number of confirmed COVID-19 cases by Italian province on March 30, 2020. Panels (c) and (d) show binscatter plots with provinces more than 50 kilometers from Lodi as the unit of observation. To generate the plot in Panel (c) we group $log(SCI)$ into 30 equal-sized bins and plot the average against the corresponding average case density. Panel (d) is constructed in a similar manner. However, we first regress $log(SCI)$ and cases per 10,000 residents on a set of control variables and plot the residualized values on each axis. Red lines show quadratic fit regressions. The controls for Panel (d) are 20 dummies for quantiles of the province's geographic distance to Lodi; GDP per inhabitant; and population density.} 
\end{minipage}
\end{figure}

Figure \ref{fig:italy} explores the analogous relationships for Lodi province in Italy.\footnote{Because Italian provinces on the island of Sardinia do not align with European NUTS3 regions (the level at which we measure social connectedness), we include Sardinia as a single observation in our analysis.} The provinces with highest COVID-19 case densities and connectedness to Lodi are in the surrounding Lombardy region, as well as the nearby Piemonte and Veneto regions. There are also relatively high levels of both connectedness to Lodi and COVID-19 cases in Rimini, a popular tourist destination along the Adriatic sea. A number of provinces in southern Italy send workers and students to the industrial Lombardy region, and therefore have strong social ties to that region. While some of these areas have seen a number of COVID-19 cases, they are not disproportionally larger, perhaps reflecting  the efforts of Italian authorities to restrict the movement of individuals \citep{kington_lat}. Panels (c) and (d) repeat the binscatter exercises from Figure \ref{fig:usa} (there are fewer data points in Figure \ref{fig:italy} than there are in Figure \ref{fig:usa}, since there are fewer Italian provinces than U.S. counties). We exclude provinces within 50 kilometers of Lodi. In Panel (d) we control for geographic distance using 20 dummies for quantiles of distance from each province to Lodi, as well as GDP per inhabitant and population density. As before, we find that the \textit{Social Connectedness Index} appears to have predictive power above these other measures that might commonly be used to proxy for social interactions. Quantitatively, the estimates from Panel (d) suggest that a doubling of the $SCI$ corresponds to an increase of 16.6 COVID-19 cases per 10,000 residents. The incremental R-squared of including social connectedness to Lodi over the other control variables is 0.057.\footnote{In Appendix \ref{sec:appendix_hotspot}, we conduct an additional exercise to mimic a potential real-world use case in which U.S. public health officials might have seeked to predict disease spread from the initial Westchester outbreak. Since, by March 10, there was not yet enough documented domestic COVID-19 spread to parameterize a forcecasting model based on U.S. data, these officials might have looked to Italy to understand how social connections to early hotspots translate into subsequent case growth. To replicate such an analysis, we train a model using Italian provinces and their conntectedness to Lodi to predict Italian COVID-19 cases as of March 10. We then use that model to forecast U.S. COVID-19 cases as of March 30 based on counties' social connectedness to Westchester as the initial hotspot. We find that including social connectedness as a model input in this exercise improves out-of-sample predictions of COVID-19 cases across U.S. counties.}

Taken together these case studies illustrate the potential usefulness of our measure of social connectedness for predicting disease spread. In the next section, we will use a time series of case growth from March through November, as well as additional predictive measures from smartphone locations and Google searches, to explore this potential in more detail.

\section{Time Series Analysis} \label{sec:series}

In this section, we exploit the changing geography of the pandemic in the U.S. to more systematically investigate the predictive value of the $Social \ Connectedness \ Index$ for forecasting the spread of COVID-19. We construct two primary time-varying metrics: ``Social Proximity to Cases'', a county-level measure of exposure to COVID-19 cases through social networks, and ``Physical Proximity to Cases'', a county-level measure of exposure through physical proximity. While the two measures will be related \citep[because individuals generally have stronger social ties to those who are geographically nearby, as documented in][]{Bailey2018_measure}, the examples in the previous section illustrate that some geographically distant places --- such as Westchester and the east coast of Florida --- can have strong social ties. To benchmark the predictive power of social connectedness, we also construct measures using data from smartphone locations and Google symptom searches.

\vspace{-0.2cm}\paragraph{Key Variable Construction.} We construct our measure of social proximity to cases as:
\begin{equation}
	\label{eq:soc_prox}
	Social \ Proximity \ to \ Cases_{i,t} = \sum_j{Cases \ Per \ 10k_{j,t} * \frac{Social \ Connectedness_{i,j}}{\sum_h{Social \ Connectedness_{i,h}}}}.
\end{equation}
$Cases \ Per \ 10k_{j,t}$ is the number of confirmed COVID-19 cases per 10,000 residents in county $j$ as of time $t$. The sums $j$ and $h$ are over all counties. Analogously, we construct a measure of a county's physical proximity to cases as:
\begin{equation}
	\label{eq:phys_prox}
	Physical \ Proximity \ to \ Cases_{i,t} = \sum_j{Cases \ Per \ 10k_{j,t}  * \frac{1}{1 + Distance_{i,j}}}.
\end{equation}
Here, $Distance_{i,j}$ is the physical distance between counties $i$ and $j$ measured in miles. We create a further related exposure measure using smartphone location data. Specifically, \cite{couture2020measuring} create a Location Exposure Index (LEX) that measures, among smartphones that pinged in a given county $i$ today, the share that pinged in each county $j$ at least once during the previous 14 days. We use these matricies to construct:
\begin{equation}
	\label{eq:lex_prox}
	LEX \ Proximity \ to \ Cases_{i,t} = \sum_j{Cases \ Per \ 10k_{j,t} * \frac{LEX_{i,j,t}}{\sum_h{LEX_{i,h,t}}}}.
\end{equation}

We also use data from \cite{google_symptoms} on searches related to COVID-19 symptoms. The data include a county by week normalized (within county) probability that a user will make a symptom-related search. For each county and two-week period, we define the change in searches related to a symptom as the percent change in this probability between the second week of the period and the second week of the previous period. We use searches related to fever, cough, and fatigue. We provide additional details in Appendix \ref{sec:appendix_google}. Finally, to explore whether it is the specific bilateral patterns of connectedness or a county's overall level of connectedness that is most relevant for predicting the spread of COVID-19, we include controls for the share of a county's Facebook connections that are within 50 and 150 miles.\footnote{With the underlying assumption that Facebook usage rates (as a share of the true population) are roughly equal across counties, we construct these controls using the \emph{Social Connectedness Index} as $Share \ Within \ K \ MI_{i} $ = $[SCI_{i,j}*Pop_{j}*1(Distance_{i,j}<K)] / [\sum_{h}{SCI_{i,h}*Pop_{h}}]$, for county $i$.}

\vspace{-0.1cm}\paragraph{Empirical Specification.}  We first study the relationship between observed case growth and ``lagged'' (i.e., in past time periods) growth in our measures. We hypothesize that if social connectedness is an important predictor of the path of COVID-19 spread, a lagged measure of social proximity to new cases will have a positive relationship with new case counts in the next period. For each county $i$ and time period $t$, our baseline specification is:
\begin{eqnarray}
	\label{eq:growth}
	log(\Delta \ Cases \ per \ 10k + 1)_{i,t} &=& 
           \beta_{1}*log(\Delta Cases \ per \ 10k + 1)_{i,t-1} \nonumber \\ 
	& + &   \beta_{2}*log(\Delta Cases \ per \ 10k + 1)_{i,t-2} \nonumber \\
	& + &    \beta_{3}*log(\Delta Social \ Proximity \ to \ Cases)_{i,t-1} \nonumber \\ 
	& + &  \beta_{4}*log(\Delta Social \ Proximity \ to \ Cases)_{i,t-2} \nonumber \\
	& + &  \beta_{5}*Share \ Friends \ within \ 50mi_i  \nonumber \\
	& + &  \beta_{6}*Share \ Friends \ within \ 150mi_i  \nonumber \\
	& + &    \beta_{7}*log(\Delta Physical \ Proximity \ to \ Cases)_{i,t-1} \nonumber \\ 
	& + &  \beta_{8}*log(\Delta Physical \ Proximity \ to \ Cases)_{i,t-2} \nonumber \\
	& + &  X_{i,t}  + \epsilon_{i,t}
\end{eqnarray}
Here, $t$ is defined as one of the eight two-week time periods between March 30 and November 2, 2020. For each time period $t$, prior two-week periods are denoted $t-2$ and $t-1$ (for example, March 3 - 16 and March 16 - 30 for the first period starting March 30). We always include two lags of own case growth, and explore the effects of lagged changes of social and physical proximity to cases. In some specifications we will add controls for $log(\Delta LEX \ Proximity \ to \ Cases + 1)_{i,t}$, lagged by one and two time periods, and for the percent change in Google searches related to fever, cough, and fatigue for this period or lagged by one period. $X_{i,t}$ are a set of time-specific fixed effects, including percentiles of population density and median household income. In our strictest specification, we also add time $\times$ state fixed effects. To rule out differential testing across regions driving our results, we also conduct a similar exercise replacing COVID-19 cases with COVID-19-related deaths. For these analyses, we use four-week time periods, beginning with April 28 - May 25, with our exposure measures lagged by four and eight weeks.

\begin{table}[hp]
    \caption{COVID-19 Case Growth and Prior Proximity to Cases} 
    \label{tab:growth_baseline}
    \includegraphics[scale = 0.95, clip, trim = 1.95cm 5.9cm 2.1cm 2.8cm]{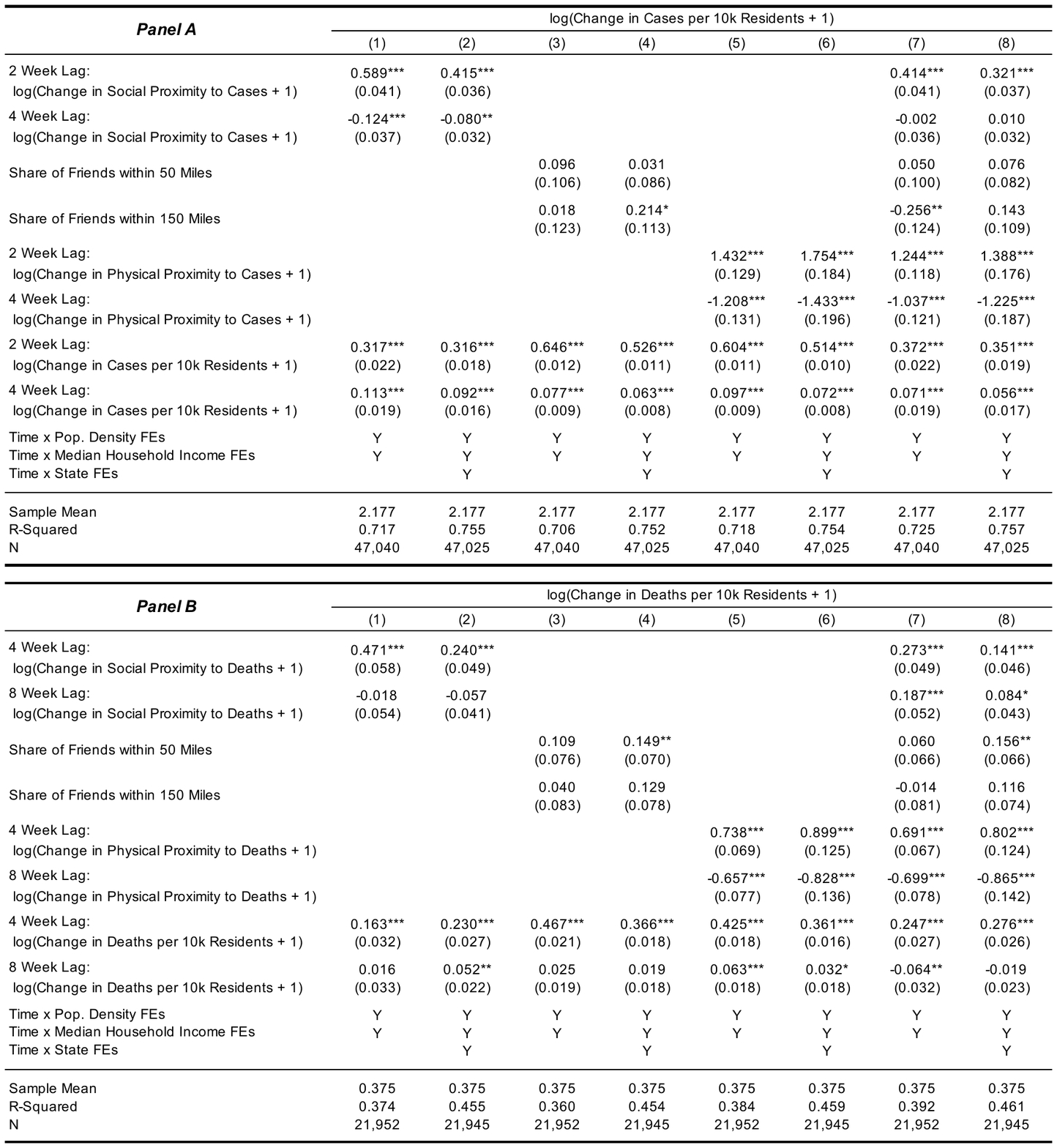}
     \begin{minipage}{\textwidth} \setstretch{.9} \medskip
          \footnotesize{{\bf Note:} Table shows results from regression \ref{eq:growth}. In Panel A, each observation is a county $\times$ two-week period (between March 30 and November 2, 2020). The dependent variable is log of one plus the number of new COVID-19 cases per 10,000 residents. In Panel B, each observation is a county $\times$ four-week period (between April 28 and November 2, 2020). The dependent variable is log of one plus the number of new COVID-19 deaths per 10,000 residents. Columns 1 and 2 include log of growth in social proximity to cases (deaths) lagged by one and two periods (two and four weeks in Panel A, four and eight weeks in Panel B). Columns 5 and 6 include analogous measures of physical proximity to cases (deaths). Columns 3 and 4 also control for the share of a county's Facebook connections that are within 50 and 150 miles. Columns 7 and 8 include all measures. All columns include controls for one and two period lagged changes in cases (deaths), as well as time-specific fixed effects for percentiles of county population density and median household income. Columns 2, 4, 6, and 8 include additional time $\times$ state fixed effects. Standard errors are clustered at the time $\times$ state level. Significance levels: *(p$<$0.10), **(p$<$0.05), ***(p$<$0.01).} 
     \end{minipage}
\end{table}

\vspace{-0.2cm}\paragraph{Regression Analysis.} Panel A of Table \ref{tab:growth_baseline} shows that past growth in social proximity to COVID-19 cases in one period has a strong positive relationship with actual growth in cases in the subsequent period. In columns 1 and 2, we document this relationship without controlling for physical distance to cases (column 2 adds state $\times$ time fixed effects to the specification in column 1, to control for time-varying state-level differences in public health measures). In contrast, columns 3 and 4 show that there is no systematic relationship between the share of a county's friends that are within 50 and 150 miles and COVID-19 cases. This suggests that it is the specific bilateral patterns of social connections that correlate with disease spread, not simply that counties with more ``open'' networks experience worse outbreaks in every period. Columns 5 and 6 show that physical proximity to cases is also strongly correlated with subsequent case growth, a relationship which may confound the one in columns 1 and 2. To address this, columns 7 and 8 include both the physical proximity and social proximity measures. While the coefficient on social proximity to cases declines somewhat --- suggesting some of the relationship is due to physical proximity --- the relationship remains highly statistically and economically significant. In our strictest specification, which includes state $\times$ period fixed effects, a doubling of social proximity to cases in one period corresponds to a 24.9\% increase in cases per capita in the next period.

Panel B of Table \ref{tab:growth_baseline} presents the same specifications, using COVID-19 deaths (instead of COVID-19 cases) as the dependent variable. The relationships are very similar, suggesting that our results are not driven by differential testing across counties that might have been correlated with social proximity to cases.

In Appendix \ref{sec:appendix_time_series} we conduct two additional regression exercises. First, we run regression \ref{eq:growth} separately for each time period, allowing us to study how the relationship between social connections and new COVID-19 cases changes over the course of the pandemic. Table \ref{tab:growth_by_period} shows that, in every two-week period from March 30 to November 2, a one-period lagged measure of social proximity to cases was a statistically significant predictor of actual case growth. In Table \ref{tab:growth_other}, we add additional measures from smartphone locations and symptom searches to our regression framework. We find that changes in Google symptom searches --- both in the current period and in the previous period --- and lagged LEX proximity to cases are strongly correlated with present case growth. However, even in the presence of each of these other predictors, changes in the social proximity to cases remains a significant predictor of subsequent case growth in sample. We next benchmark the predictive power of social connectedness to these measures using an out-of-sample prediction exercise.

\vspace{-0.2cm}\paragraph{Out-of-Sample Prediction Analysis.}

Building on our previous results, we next conduct a simple out-of-sample prediction exercise. During a pandemic, local policymakers might want to determine their localities' risks for an outbreak in real time to inform public health measures. With this case in mind, we build a series of simple models that use available data at time $t$ to predict case growth in counties at time $t+1$. We test the added predictive value of social proximity to cases by building separate models that include and exclude this measure, as well as other possible predictors. Because we do not use the ``test'' data to train the models, a reduction in prediction error would be reflective of a true improvement in real-world predictions of COVID-19 cases that could have been achieved from using social connectedness data 	(as opposed to the increase in in-sample $R^2$ in our previous analyses).

\begin{table}[hp]
    \caption{Predicting COVID-19 cases in U.S., with and without Social Proximity to Cases} 
    \label{tab:rmse}
    \includegraphics[scale = 0.95, clip, trim = 1.9cm 15.4cm 2.1cm 2.5cm]{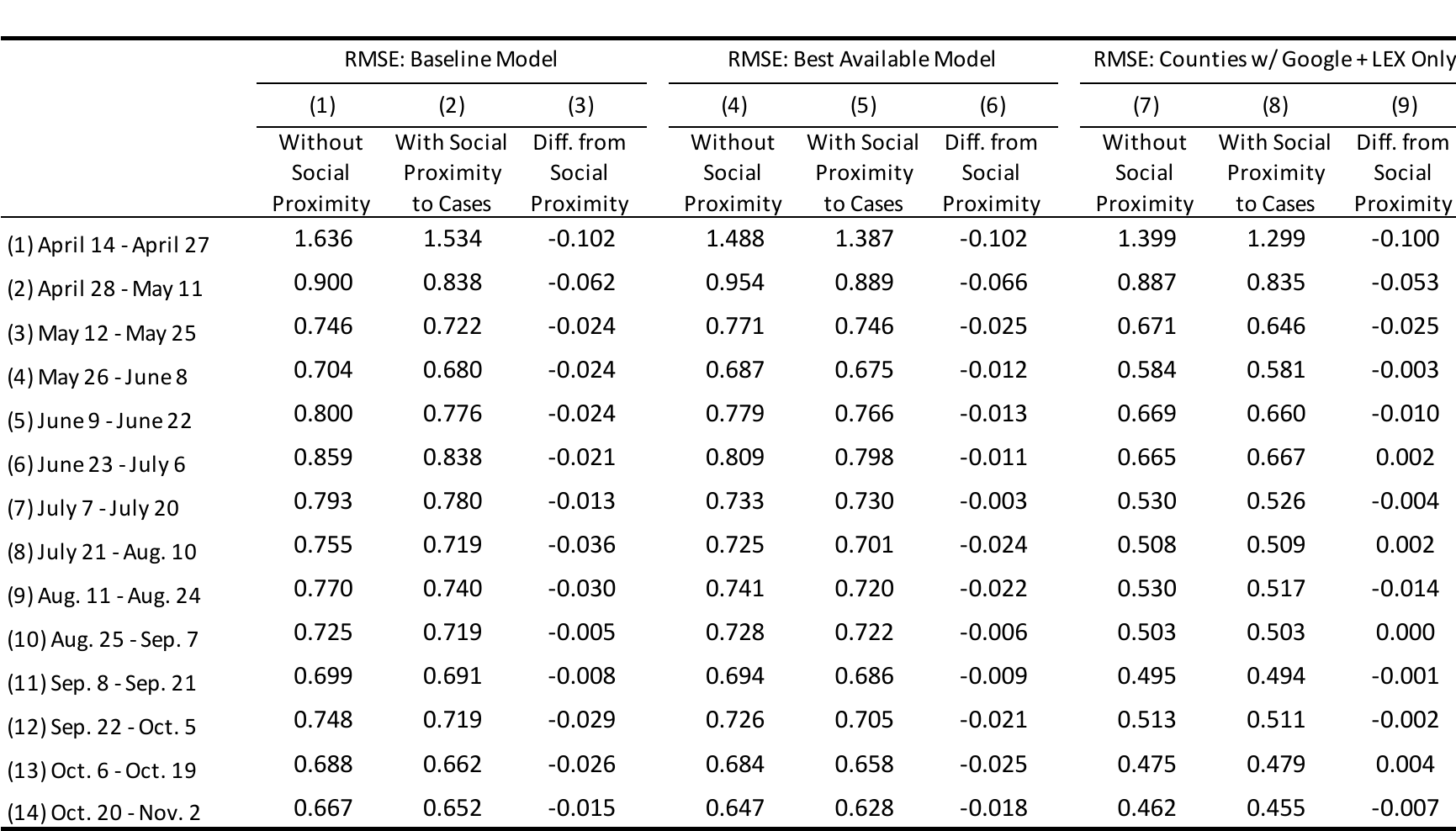}
     \begin{minipage}{\textwidth} \setstretch{.9} \medskip
          \footnotesize{{\bf Note:} Table shows results from county-level predictions of COVID-19 case growth. The predicted outcome is log of one plus the number of new COVID-19 cases per 10,000 residents. All columns show root mean squared errors (RMSEs) from a random forest model trained on data from all periods prior to the period of interest. The model inputs in column 1 are population density; median household income; and log of growth in physical proximity to cases and actual cases, lagged by two and four weeks (one and two time periods). Columns 4 and 7 include information on one and two period lagged measures of LEX proximity to cases, and one period lagged percent changes in Google searches related to fever, cough, and fatigue. Column 4 includes predictions for 3,136 counties using, for each county, a model that utilizes the most available information possible. Column 7 limits to the 1,976 counties for which we have both Google symptom search and LEX data. Columns 2, 5, and 8 add lagged measures of social proximity to cases to columns 1, 4, and 7. Columns 3, 6, and 9 show the change in RMSE from adding social proximity to cases.} 
     \end{minipage}
\end{table}

Table \ref{tab:rmse} shows the results of this prediction exercise. The results in all columns are generated using a random forest, an ensemble prediction algorithm commonly used in data science applications. The algorithm allows us to find non-linear relationships between variables, without overfitting, by aggregating mean predictions from a number of regression trees generated over sample subsets of both observations and input variables.\footnote{In our analysis we use 500 trees. For more information on random forests, see \cite{breiman2001random}.}  In most settings, random forest out-of-sample predictions outperform those of linear models.

Columns 1-3 describe the prediction error from a simple model that includes the measures from columns 7 and 8 in Panel A of Table \ref{tab:growth_baseline}.\footnote{Here, we use non-binned measures of population density and median household income. Because random forests are able to identify non-linear relationships between input and outcome measures, we no longer need to split the measures into percentiles to allow for non-linear relationships.} Column 1 excludes the two lagged measures of social proximity to cases and column 2 includes them. Columns 1 and 2 show the root mean squared error (RMSE) from a model trained using data from all periods before the period of interest, then tested on that next period; each prediction period is shown as a separate row. The RMSE for both models generally decreases as the training sample gets larger, ending at 0.667 and 0.652 log new cases per 10,000 residents. Column 3 shows the difference in RMSE between the two models, with negative numbers indicating an improvement in out-of-sample fit from including social proximity to cases. In every row, the RMSE is lower when including social proximity to cases, suggesting it does significantly improve predictions.

In columns 4-9 we add information on Google symptom searches and mobility based on smartphone locations. Doing so allows us to benchmark the predictive value of social proximity to cases over and above these other predictors. Columns 4-6 include predictions for all counties included in the COVID-19 case data. We make a prediction for each county using the ``best'' model (in terms of model features) based on data availability. For example, for a county with LEX and Google data, we will predict cases using a model trained with LEX and Google data. For a county with only LEX data, we will predict using a model trained without Google data, and so on.\footnote{This requires training four models: (1) baseline; (2) baseline + LEX; (3) baseline + Google; and (4) baseline + LEX + Google. We train each model on every county that has the necessary non-missing data. For example, a county with LEX and Google data will be used to train all four models.} Column 6 shows that, once again, RMSE decreases in every period after including social proximity to cases, highlighting its incremental predictive value over and above other measures one might have used. 

In columns 7-9 we limit to the 1,976 counties which have both LEX and Google symptom search data. Column 9 shows that in 10 of 14 periods, predictions using social connectedness do outperform the comparison model. However, the differences are generally small, suggesting that when limiting to \emph{only} counties with LEX and Google data, social proximity to cases may provide only a small degree of additional predictive value. This is perhaps unsurprising: our proposed mechanism by which social connectedness helps forecast COVID-19 spread is through predicting in-person interactions, which are more directly measured in LEX data.\footnote{In Appendix \ref{sec:appendix_deaths} we repeat this exercise using COVID-19 related deaths. We find similar results. Baseline models that include social proximity to deaths have smaller prediction errors than models that exclude the measure. Social proximity to deaths does not appear to add sizable predictive value to LEX and Google measures when limiting to counties with both sets of data; however, when using the ``best available'' model in any county, social proximity to deaths \emph{does} sizably improve predictions.} 

Tthe fact that social connectedness consistently improves predictions in the full set of U.S. counties (columns 4-6) highlights an important availability advantage of the data. While the LEX and Google data are limited to counties with a sufficient number of devices or searches in a period, the relatively stable nature of social connectedness over time (combined with Facebook's large user base) allows the $SCI$ to be available in more counties, and potentially also at finer levels, such as zip codes.\footnote{The SCI data used for this paper include 3,141 counties, the Google data include 2,572 counties, and the LEX data include 2,018 counties.} Furthermore, Facebook's global reach allows for $SCI$ measures \emph{within and between} most parts of the world. We are unaware of smartphone location data that can similarly measure, for example, connections between GADM1 regions in Africa, NUTS3 regions in Europe, and U.S. counties --- and information on these connections may aid in forecasting the global spread of communicable diseases.

\section{Conclusion} \label{sec:discussion}

In the context of threats from communicable diseases such as COVID-19, a region's ability to determine optimal public health responses depends on its ability to forecast the risk of an outbreak \citep{reich2019collaborative}. A primary determinant of this risk is the likelihood of physical interactions between the region's residents and residents of other areas with severe outbreaks. Information on the geography of social connections, which shape patterns of physical interactions, are therefore crucially important for public health officials. In this paper, we use de-identified and aggregated data from Facebook to measure social connections between regions, and find those connections to be an important predictor of outbreaks during the COVID-19 pandemic. We show that areas that are more connected to early pandemic hotspots in the U.S. and Italy had, on average, higher case counts by March 30, 2020, even after controlling for physical distance and other demographics. Furthermore, due to its broad geographic coverage, social connectedness data improves out-of-sample predictions of COVID-19 spread during the U.S. pandemic beyond smartphone location and Google symptom search data. 

The methodologies we use should not be interpreted as an attempt to create a state-of-the-art epidemiological model. However, our results strongly suggest that our measure of social connectedness may prove useful in future epidemiological work. In particular, its high-degree of availability --- in terms of both geographic coverage and granularity --- allow social connectedness to provide predictive power over and above other available measures.

\bibliography{bib}

\clearpage

\renewcommand{\thefigure}{A\arabic{figure}}
\setcounter{figure}{0}

\renewcommand{\thetable}{A\arabic{table}}
\setcounter{table}{0}

\begin{appendices}

\section{Out-of-Sample Hotspot Analysis}\label{sec:appendix_hotspot}

In this section, we conduct a simple prediction exercise to test the value of social connectedness in forecasting how a disease will propagate from a hotspot early in a pandemic. To do so, we train a model using an Italian hotspot and test it using a subsequent U.S. hotspot.

On March 10, 2020, New York state created a ``containment area'' around New Rochelle, a community in Westchester County that had the first major COVID-19 outbreak in the eastern United States. At this time, U.S. local officials around the country were likely worried about the extent of their own regions' exposures to COVID-19. Yet, in the absence of existing data on domestic COVID-19 spread in the U.S., it may have been difficult to calibrate a local forecasting model. One potential solution would have been to train a model using information from the Italian outbreak --- which by mid-March had been spreading for some time --- to predict COVID-19 spread in the U.S.

We mimic this use case by first training linear regression and random forest models using data from Italy. Specifically, for each Italian province, we predict the number COVID-19 cases per 10,000 people on March 10 using population density, a measure of income, and distance and social connectedness to the Lodi hotspot. We exclude provinces within 80 km (50 miles) of Lodi. We train versions of these models with and without including social connectedness to Lodi as a predictor. In each model, we normalize every measure by subtracting the mean and dividing by the standard deviation. This normalization, which is common in machine learning prediction applications, ensures that our predictive measures are scaled similarly in the U.S. and Italian settings. In our prediction exercise we use, for each U.S. county, the Italy-trained models to predict the number of COVID-19 cases per 10,000 people on March 30 (i.e., 20 days in the future) with and without data on social connectedness to Westchester. 

\begin{table}[h]
    \caption{Predicting U.S. Hotspot COVID-19 spread, trained on Italian Hotspot spread} 
    \label{tab:hotspot_predictions}
    \includegraphics[scale = 0.95, clip, trim = 1.95cm 22cm 2.1cm 2.8cm]{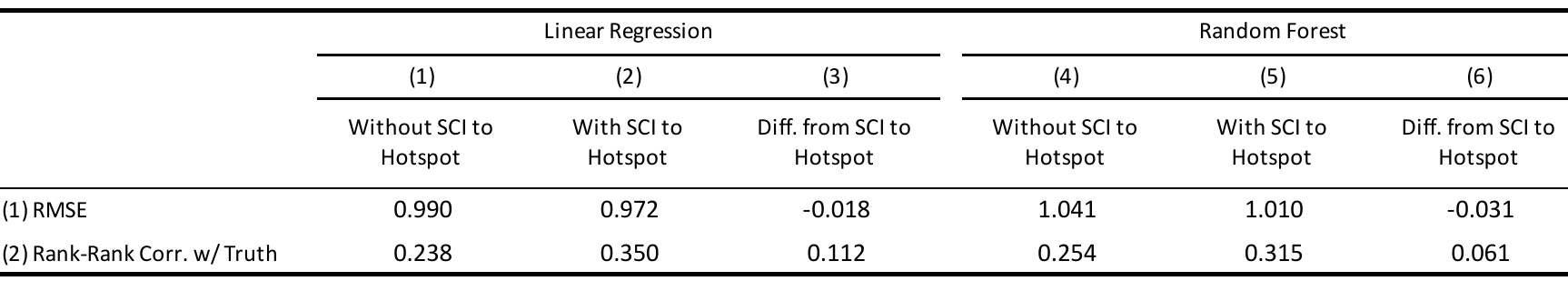}
     \begin{minipage}{\textwidth} \setstretch{.9} \medskip
          \footnotesize{{\bf Note:} Table shows results from county-level predictions of COVID-19 cases per 10,000 residents. Columns 1-3 and 4-6 show results from linear regression and random forest models, respectively. The models are trained using information from Italy on March 10 and tested using information from the U.S. on March 30. All measures are normalized by subtracting the mean then dividing by the standard deviation. Row (1) shows the prediction root mean squared errors (RMSEs) and row (2) shows prediction rank-rank correlation with the truth. The model inputs in columns 1 and 4 are  $log(distance)$ to the hotspot (Lodi in Italy, Westchester in the U.S.), population density, and household income / (GDP per inhabitant). Columns 2 and 5 add $log(SCI)$ to the hotspots. Columns 3 and 6 show the change in each measure from adding $log(SCI)$.}
     \end{minipage}
\end{table}

Table \ref{tab:hotspot_predictions} shows that our predictions improve when adding social connectedness to the hotspots as a model input. Row (1) shows that the RMSE drops from 0.99 to 0.97 (in standard deviations from the mean cases per 10,000 residents) for the linear regression model and from 1.04 to 1.01 for the random forest model. Row (2) compares the rank of counties' predictions with the true rank. For both models the rank-rank correlations increase when including social connectedness to the hotspot as a predictor.

\section{Additional Time Series Regressions}\label{sec:appendix_time_series}

Table \ref{tab:growth_by_period} shows the results from running the regression in Table \ref{tab:growth_baseline}'s column 8 separately for every two-week period betwen March 30 and November 2. In every period, a two week lagged measure of social proximity to cases was a statistically significant predictor of actual case growth.  The magnitudes of the coefficients suggest a doubling in social proximity to cases in one two-week period corresponds to between a 10.9\% and 66.4\% increase in actual cases in the next period, after controlling for physical proximity to cases and other controls.

Table \ref{tab:growth_other} shows the results from adding additional predictive measures to Table \ref{tab:growth_baseline}. Columns 1 and 2 are the same as columns 7 and 8 in Table \ref{tab:growth_baseline}. Columns 3 and 4 show that a ``nowcast'' of changes in Google symptom search trends is strongly correlated with changes in case growth. Columns 5 and 6 show that this relationship persists, though somewhat less strongly, for a one-period lagged measure of changes in symptom searches. The latter measure, unlike the former, could be used to predict \emph{future} case growth. Columns 7 and 8 show that the change in LEX proximity to cases in one period also has a strong positive relationship with actual case growth in the next. Even in the presence of each of these other measures, social proximity to cases remains a significant predictor of future case growth. 

\begin{landscape}
\begin{table}[p]
    \caption{COVID-19 Case Growth and Prior Proximity to Cases, by Two-Week Period} 
    \label{tab:growth_by_period}
    \includegraphics[scale = 1.3, clip, trim = 1.9cm 18.8cm 1cm 2.6cm]{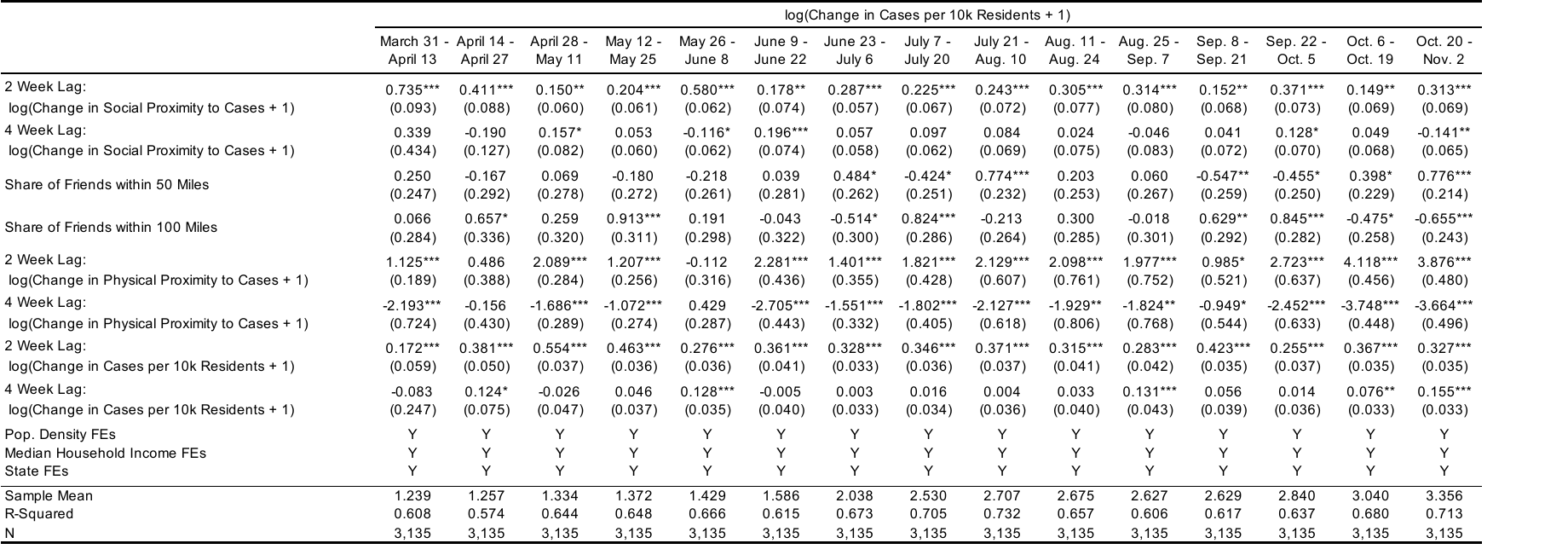}
     \begin{minipage}{\linewidth} \setstretch{1} \medskip
          \footnotesize{{\bf Note:} Table shows time-specific results from regression \ref{eq:growth}. Each observation is a county. The dependent variable is log of one plus the number of new COVID-19 cases per 10,000 residents in one two-week period between March 30 and November 2, 2020. All columns include log of growth in social and physical proximity to cases, as well as log of growth in actual cases, lagged by two and four weeks (one and two time periods). All columns include time-specific fixed effects for percentiles of population density and median household income, time-specific fixed effects for state, and estimations of the share of a county's Facebook connections that are within 50 and 150 miles  Significance levels: *(p$<$0.10), **(p$<$0.05), ***(p$<$0.01).} 
     \end{minipage}
\end{table}
\end{landscape}

\begin{table}[hp!]
    \caption{COVID-19 Case Growth, Prior Proximity to Cases, and Other Predictive Measures} 
    \label{tab:growth_other}
    \includegraphics[scale = 0.95, clip, trim = 1.95cm 9.3cm 2.1cm 2.8cm]{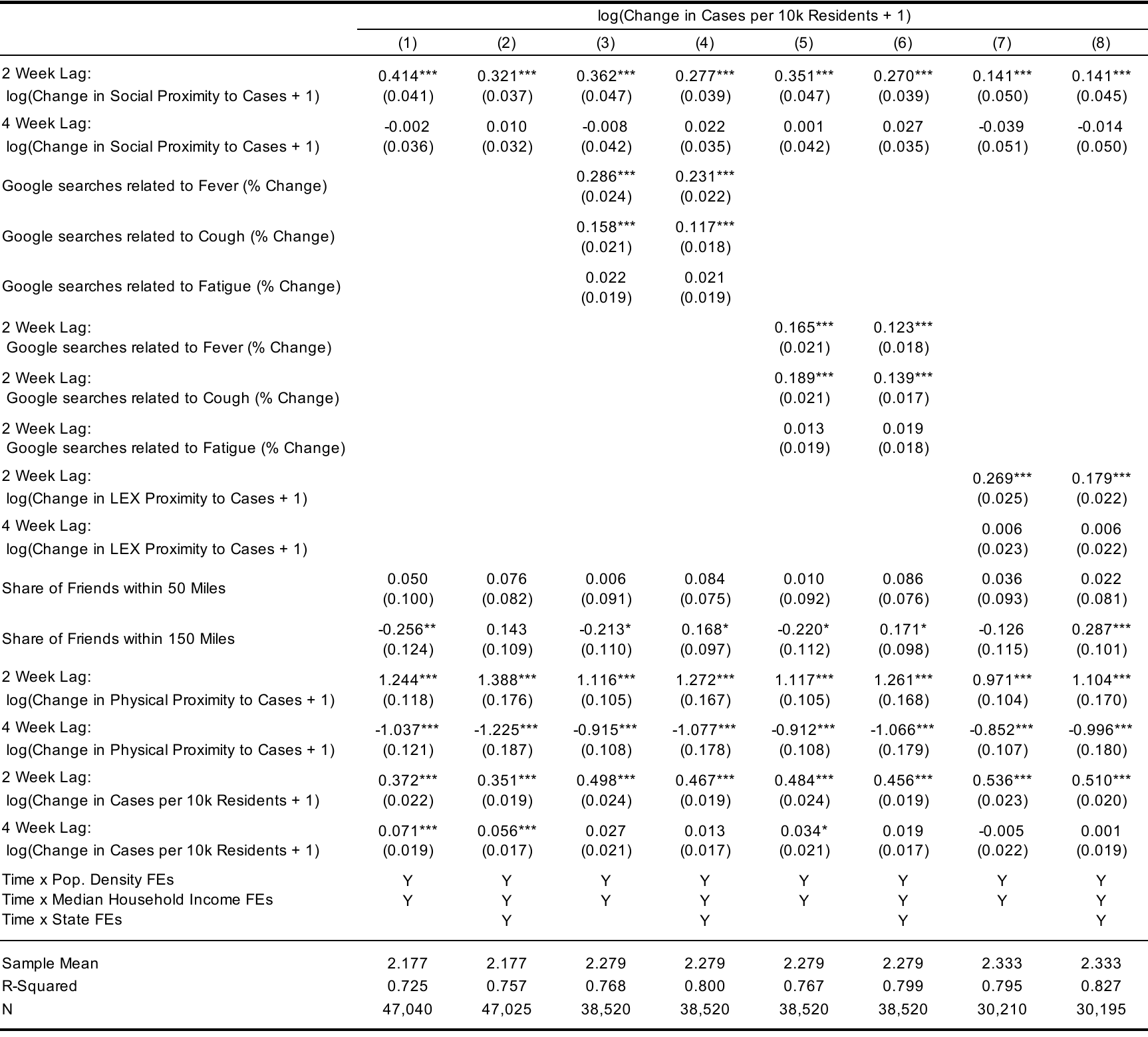}
     \begin{minipage}{\textwidth} \setstretch{.9} \medskip
          \footnotesize{{\bf Note:} Table shows results from regression \ref{eq:growth}. Each observation is a county $\times$ two-week period (between March 30 and November 2, 2020). The dependent variable is log of one plus the number of new COVID-19 cases per 10,000 residents. Columns 1 and 2 are the same as columns 7 and 8 in Table \ref{tab:growth_baseline}. Columns 3 and 4 add the percent growth in Google searches related to fever, cough, and fatigue from the week prior to the period to the second week of the period. Columns 5 and 6 includes analogous measures lagged by one period. Columns 7 and 8 add LEX-based proximity to cases. Standard errors are clustered at the time $\times$ state level. Significance levels: *(p$<$0.10), **(p$<$0.05), ***(p$<$0.01).}
     \end{minipage}
\end{table}

\section{Out-Of-Sample Prediction: COVID-19 Deaths}\label{sec:appendix_deaths}

Table \ref{tab:rmse_deaths} show the prediction error from a simple out-of-sample forecasting exercise analogous to that in Table \ref{tab:rmse}, but for COVID-19 deaths instead of COVID-19 cases. Columns 1-3 use the variables in columns 7 and 8 of Table \ref{tab:growth_baseline}, Panel B as predictors. The model presented in column 2 includes the two lagged measures of social proximity to COVID-19 deaths, while the model presented in column 1 does not include them. The RMSE is lower in every period in the model that includes social proximity to deaths as a predictor. Columns 4-6 include predictions for all counties included in the COVID-19 case data. We make a prediction using the ``best'' model available (in terms of number of model features), as described in Section \ref{sec:series}. Column 6 shows that the RMSE decreases again in every period when including social proximity to deaths as a predictor. Columns 7-9 limit to counties with Google and LEX data. Similar to our findings for county-level COVID-19 cases, social connectedness does not appear to provide substantial additional predictive value in this particular setting.

\begin{table}[htb]
    \caption{Predicting COVID-19 deaths in U.S., with and without Social Proximity to Deaths} 
    \label{tab:rmse_deaths}
    \includegraphics[scale = 0.95, clip, trim = 1.9cm 19.6cm 2.1cm 2.5cm]{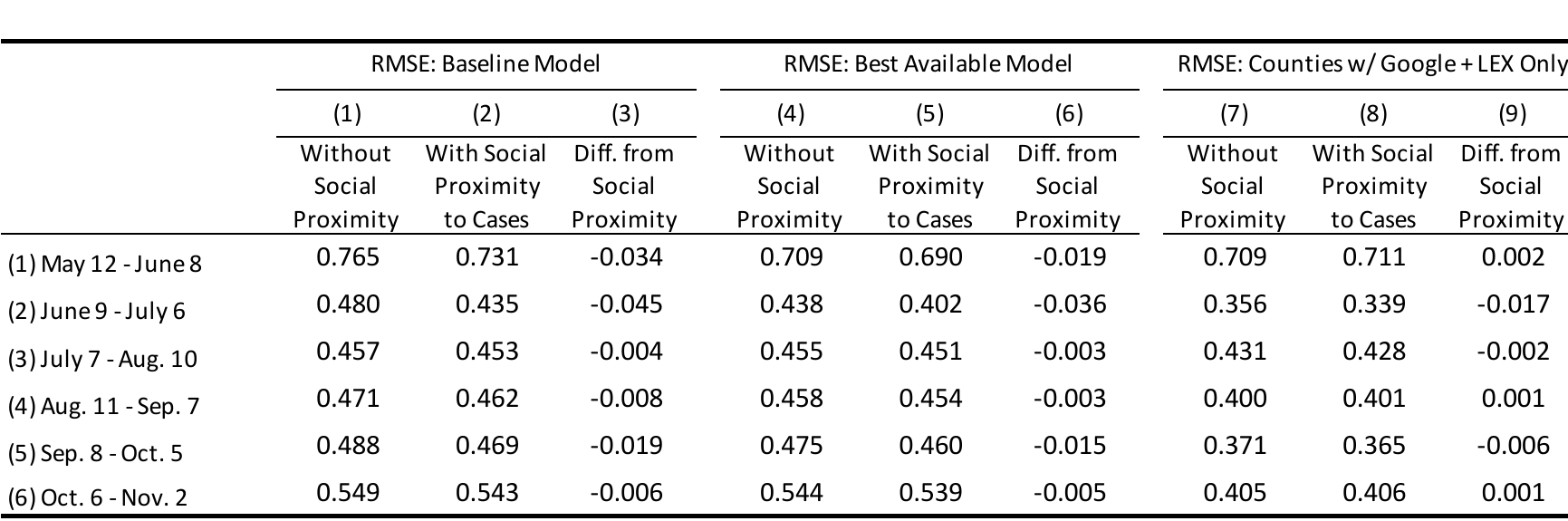}
     \begin{minipage}{\textwidth} \setstretch{.9} \medskip
          \footnotesize{{\bf Note:} Table shows results from county-level predictions of COVID-19 deaths. The predicted outcome is log of one plus the number of new COVID-19 deaths per 10,000 residents. All columns show root mean squared errors (RMSEs) from a random forest model trained on data from all periods prior to the period of interest. The model inputs in column 1 are population density; median household income; and log of growth in physical proximity to deaths and actual deaths, lagged by four and eight weeks (one and two time periods). Columns 4 and 7 include information on one and two period lagged measures of LEX proximity to deaths, and one period lagged percent changes in Google searches related to fever, cough, and fatigue. Column 4 includes predictions for 3,156 counties using, for each county, a model that utilizes the most available information possible. Column 7 limits to 1,976 counties for which we have both Google symptom search and LEX data. Columns 2, 5, and 8 add lagged measures of social proximity to deaths to columns 1, 4, and 7. Columns 3, 6, and 9 show the change in RMSE from adding social proximity to deaths.} 
     \end{minipage}
\end{table}

\section{Additional Details on Google Symptom Search Data}\label{sec:appendix_google}

In Section \ref{sec:series}, Appendix \ref{sec:appendix_time_series}, and Appendix \ref{sec:appendix_deaths} we use data on Google searches related to COVID-19 symptoms from \cite{google_symptoms}. The data include a county by week normalized (within county) probability a user will make a symptom-related search. These measures come in a daily series and a weekly series. If a given symptom in a given county does not meet certain Google quality or privacy thresholds at the daily level, it will be provided at the weekly level. If it cannot meet these thresholds at the weekly level, it is not provided. If a county/symptom is provided for a period at the daily level, it is not also provided at the weekly level. To create a time series at the county/weekly level, we therefore average non-missing daily measures by week. We use searches related to three common COVID-19 symptoms: fever, cough, and fatigue.

To aggregate our measure to the bi-weekly periods used in Section \ref{sec:series}, Appendix \ref{sec:appendix_time_series}, and Appendix \ref{sec:appendix_deaths} we define the change in searches related to a symptom as the percent change in the probability between the second week of the period and the second week of the previous period. For example, for the period March 31 - April 13, the percent change is from March 23 - March 29 to April 6 - 12. In our prediction exercise presented in Table \ref{tab:rmse}, we use a one-period lagged version of this measure. Some counties are missing data for particular weeks. So that our final sample has a complete time series for every included county, we exclude counties with missing data in more than half the periods in our sample and impute zero change in any remaining missing periods. 97 percent of counties included are missing fewer than 1 in 4 periods and 77 percent are missing fewer than 1 in 20 periods.

\end{appendices}

\end{document}